\documentclass[lettersize,journal]{IEEEtran}
\IEEEoverridecommandlockouts
\usepackage{cite}
\usepackage{amsmath,amssymb,amsfonts,bbm,bm}
\usepackage{algorithm}
\usepackage{algorithmic}
\usepackage{graphicx}
\usepackage{float}
\usepackage{subfigure}
\usepackage{textcomp}
\usepackage{xcolor}
\usepackage{float}
\usepackage{paralist}
\usepackage{multirow}
\usepackage{booktabs}
\usepackage[font=small]{caption}
\usepackage{subcaption}
\usepackage{graphicx}
\usepackage{tikz}
\usetikzlibrary{shapes.geometric, arrows, positioning}
\usepackage[nolist]{acronym}
\usepackage{epstopdf}
\usepackage{dblfloatfix}

\usepackage{titlesec}
\titlespacing*{\section}{0pt}{*1}{*1}
\titlespacing{\subsection}{0pt}{*1}{*1}

\renewcommand{\thesubsubsection}{\arabic{subsubsection}}

\titleformat{\subsubsection}[runin]{\itshape}{\thesubsubsection)}{1em}{}
\titlespacing*{\subsubsection}{\parindent}{0pt}{*1}

\DeclareMathOperator*{\argmax}{argmax}

\newtheorem{remark}{Remark}

\newtheorem{lemma}{Lemma}

\newcommand{\mM}{\mathbf{M}}
\newcommand{\mS}{\mathbf{S}}
\newcommand{\mE}{\mathbf{E}}

\newcommand{\mF}{\mathbf{F}}
\newcommand{\mW}{\mathbf{W}}
\newcommand{\mI}{\mathbf{I}}
\newcommand{\mH}{\mathbf{H}}
\newcommand{\mA}{\mathbf{A}}

\newcommand{\mG}{\mathbf{G}}
\newcommand{\mL}{\mathbf{L}}

\newcommand{\mQ}{\mathbf{Q}}
\newcommand{\mT}{\mathbf{T}}
\newcommand{\mU}{\mathbf{U}}
\newcommand{\mLambda}{\boldsymbol{\Lambda}}

\newcommand{\ba}{\mathbf{a}}
\newcommand{\br}{\mathbf{r}}
\newcommand{\bnu}{\boldsymbol{\nu}}
\newcommand{\bb}{\mathbf{b}}

\newcommand{\bmm}{\mathbf{m}}
\newcommand{\bv}{\mathbf{v}}
\newcommand{\bu}{\mathbf{u}}
\newcommand{\bh}{\mathbf{h}}
\newcommand{\bs}{\mathbf{s}}
\newcommand{\bx}{\mathbf{x}}
\newcommand{\by}{\mathbf{y}}
\newcommand{\bz}{\mathbf{z}}
\newcommand{\bw}{\mathbf{w}}
\newcommand{\be}{\mathbf{e}}
\newcommand{\bl}{\mathbf{l}}

\newcommand{\bq}{\mathbf{q}}
\newcommand{\bg}{\mathbf{g}}
\newcommand{\bt}{\mathbf{t}}
\newcommand{\bepsilon}{\boldsymbol{\varepsilon}}
\newcommand{\bomega}{\boldsymbol{\omega}}

\newcommand{\setC}{\mathbb{C}}
\newcommand{\setR}{\mathbb{R}}
\newcommand{\setS}{\mathbb{S}}

\newcommand{\setL}{\mathbb{L}}

\newcommand{\txtand}{\; \text{ and } \;}
\newcommand{\txtfor}{\; \text{ for } \;}

\newcommand{\R}{\mathcal{R}}
\newcommand{\Q}{\mathcal{Q}}

\newcommand{\Lag}{\mathcal{L}}

\newcommand{\sinr}{\mathrm{SINR}}

\newcommand{\trp}{\mathsf{T}}
\newcommand{\ul}{{\mathrm{u}}}
\newcommand{\dl}{{\mathrm{d}}}
\newcommand{\her}{\mathsf{H}}
\newcommand{\set}[1]{\left\lbrace #1 \right\rbrace}
\newcommand{\tr}[1]{\mathrm{tr}\left\lbrace #1 \right\rbrace}
\newcommand{\brc}[1]{\left( #1 \right)}
\newcommand{\dbc}[1]{\left[ #1 \right]}
\newcommand{\norm}[1]{\left\Vert #1 \right\Vert}
\newcommand{\abs}[1]{\left\vert #1 \right\vert}
\newcommand{\diag}[1]{\mathrm{diag} \left\lbrace #1 \right\rbrace}
\newcommand{\Ex}[1]{\mathbbm{E} \left\lbrace #1 \right\rbrace}
\newcommand{\subto}{\text{ \normalfont s.t.} \;\;}

\def\BibTeX{{\rm B\kern-.05em{\sc i\kern-.025em b}\kern-.08em
    T\kern-.1667em\lower.7ex\hbox{E}\kern-.125emX}}

\begin{document}
\begin{acronym}
\acro{ap}[AP]{access point}
\acro{mimo}[MIMO]{multiple-input multiple-output}
\acro{los}[LoS]{line-of-sight}
\acro{noma}[NOMA]{non-orthogonal multiple access}
\acro{irs}[IRS]{reconfigurable intelligent surface}
\acro{snr}[SNR]{signal to noise ratio}
\acro{sinr}[SINR]{signal to interference and noise ratio}
\acro{bcd}[BCD]{block coordinate descent}
\acro{rzf}[RZF]{regularized zero-forcing}
\acro{pas}[PASS]{\underline{p}inching-\underline{a}ntenna \underline{s}y\underline{s}tem}
\acro{fp}[FP]{fractional programming}
\acro{mrt}[MRT]{maximal-ratio transmission}
\acro{zf}[ZF]{zero-forcing}
\acro{tdd}[TDD]{time-domain duplexing}
\acro{mse}[MSE]{mean squared error}
\acro{mmse}[MMSE]{minimum mean squared error}
\acro{mf}[MF]{matched filtering}
\end{acronym}

\title{MIMO-PASS: Uplink and Downlink Transmission via MIMO Pinching-Antenna Systems}
\author{Ali Bereyhi,~\IEEEmembership{Member, IEEE},
Chongjun Ouyang,~\IEEEmembership{Member, IEEE},
Saba Asaad,~\IEEEmembership{Member, IEEE},\\
Zhiguo Ding,~\IEEEmembership{Fellow, IEEE}, and
H. Vincent Poor,~\IEEEmembership{Life Fellow, IEEE}\vspace{-8mm}
\thanks{A. Bereyhi is with Department of Electrical and Computer Engineering at University of Toronto; email: \textit{ali.bereyhi@utoronto.ca}.}%
\thanks{Chongjun Ouyang is with School of Electronic Engineering and Computer Science at Queen Mary University of London; email: \textit{c.ouyang@qmul.ac.uk}.}
\thanks{Saba Asaad is with Department of Electrical Engineering and Computer Science at York University; email: \textit{asaads@yorku.ca}.}%
\thanks{Zhiguo Ding is with School of Electrical and Electronic Engineering at University of Manchester; email: \textit{zhiguo.ding@manchester.ac.uk}.}
\thanks{H. Vincent Poor is with Department of Electrical and Computer Engineering at Princeton University; email: \textit{poor@princeton.edu}.}
}

\maketitle

\begin{abstract}
Pinching-antenna systems (PASSs) are a recent flexible-antenna technology that is realized by attaching simple components, referred to as \textit{pinching elements}, to dielectric waveguides. This work explores the potential of deploying PASS for uplink and downlink transmission in multiuser MIMO settings. For downlink PASS-aided communication, we formulate the optimal hybrid beamforming, in which the digital precoding matrix at the access point and the location of pinching elements on the waveguides are jointly optimized to maximize the achievable weighted sum-rate. Invoking fractional programming and Gauss-Seidel approach, we propose two low-complexity algorithms to iteratively update the precoding matrix and activated locations of the pinching elements. We further study uplink transmission aided by a PASS, where an iterative scheme is designed to address the underlying hybrid multiuser detection problem. We validate the proposed schemes through extensive numerical experiments. The results demonstrate that using a PASS, the throughput in both uplink and downlink is boosted significantly as compared with baseline MIMO architectures, such as massive MIMO~and classical hybrid analog-digital designs. This highlights the great potential of PASSs, making it a promising reconfigurable antenna technology for next-generation wireless systems. 
\end{abstract}


\section{Introduction}
\label{sec:intro}




\IEEEPARstart{S}{hannon's} coding theorem characterizes the maximum data rate achievable for a given transmission model, i.e., for a given channel \cite{shannon1948mathematical}. In the context of wireless communications, the transmission model is traditionally considered to be specified by environmental factors such as~path~loss,~diffraction, and scattering that are beyond human control and manipulation. This assumption has shaped the evolution of conventional wireless \ac{mimo} designs, including fully digital and hybrid analog-digital \ac{mimo} systems, where the primary focus has been on optimizing transceiver architectures to maximize throughput \cite{larsson2014massive,sohrabi2017hybrid}.

As wireless systems move to higher frequency band, such as millimeter or terahertz waves, the impact of severe path loss becomes more pronounced \cite{rappaport2013millimeter,akyildiz2022terahertz}. The high attenuation at these frequencies imposes significant challenges in maintaining reliable links over moderate to long distances. The deployment of massive \ac{mimo} has emerged as an effective solution in this respect, as large arrays can form highly directional beams that compensate for the detrimental effects of path-loss. This empirical observation, combined with theoretical advances such as channel hardening \cite{hochwald2004multiple, chen2018channel}, has challenged this conventional assumption that the data transmission model in wireless systems is inherently uncontrollable, suggesting a paradigm shift toward emerging technologies that can actively reconfigure the channel \cite{gong2020toward}. 

Although channel reconfiguration has recently gained increased attention, it has been an active research area for a considerable time \cite{cetiner2006mimo,sayeed2007maximizing,xu2014design,asaad2017asymptotic}. Primary studies considered basic approach such as \textit{antenna selection}, in which the channel is modified by dynamically choosing a subset of available antennas \cite{molisch2004mimo,li2014energy,asaad2018massive}. Inspired by this idea, several lines of work suggest reconfigurable transceiver architectures which can modify the communication link between the transmitter and receiver \cite{asaad2017asymptotic}. The results of these early studies demonstrate the potential benefits of spatial adaptability in wireless systems. Recent advancements in channel reconfiguration have led to the development of more sophisticated technologies, such as \acp{irs} \cite{wu2019towards}, fluid antennas \cite{wong2020fluid}, and movable antenna systems \cite{zhu2023movable}. In this work, we investigate a recently-proposed reconfigurable technology called \textit{\acp{pas}} to understand its potentials for efficient channel modification in wireless systems.  

\subsection{Reconfigurable Antenna Technologies}
Advances in antenna technology have led to several breakthroughs in the design of intelligent components for dynamic environment reconfiguration \cite{alexandropoulos2020reconfigurable,ozdogan2019intelligent,konca2015frequency,jornet2013graphene}. \Ac{irs} technology is an example, which can apply controllable phase-shifts to the incident electromagnetic waves \cite{gong2020toward,wu2021intelligent}. Through these tunable phase-shifts, \acp{irs} can reconfigure the end-to-end effective channel. An \ac{irs} can be seen as a component that realizes \textit{programmable} propagation environments. By properly placing \acp{irs} in a network, various favorable propagation properties, such as bypassing obstruction, improving coverage, channel hardening, and reducing multi-user interference can be realized in a real-time and cost-efficient manner \cite{alexandropoulos2020reconfigurable,wu2019intelligent,bereyhi2023channel,zheng2021double}.

%


Fluid antennas represent another technology that provides a means for intelligent reconfiguration of the wireless medium \cite{new2024tutorial,wong2020fluid,wong2023fluid-1,wong2023fluid-2,wong2023fluid-3}. Fluid antennas deploy conductive liquids, such as mercury or galinstan, encapsulated within a dielectric holder to form an adaptive illumination. By controlling the displacement, shape, and volume of the liquid conductor, fluid antennas can adjust their electrical and radiation properties in real time. This flexibility enables dynamic beam-steering and enhanced spatial diversity, making fluid antennas advantageous for compact devices with limited physical space \cite{wong2020fluid}. Despite this, these components face challenges with respect to scalability, due to physical limitation in the implementation and control of liquid-based illuminating elements on large scales \cite{wong2020performance}. To address this issue, the \textit{movable antenna} technology has been introduced, which realizes the reconfigurable properties of fluid antennas by mechanical means \cite{zhu2023movable}.

In movable antennas, the location of the antenna elements can be dynamically tuned within a designated region of the transceiver array. Through this tuning, these antenna arrays can reconfigure the propagation environment \cite{zhu2023movable,zhu2023movable-2}. Unlike fluid antennas, the movement is controlled mechanically in this technology, enabling it to be scaled more efficiently \cite{zhu2025surveryMov}. From an abstract viewpoint, the movement of antennas in these arrays can be seen as antenna selection with a large \textit{virtual} array of passive antennas, whose elements are located at the possible positions on moving antennas \cite{zhu2023modeling,ma2023mimo}. The large array in this case is however realized implicitly via mechanical displacements. This allows for efficient implementation by bypassing the significant loss introduced by switching in the \textit{direct} implementation of antenna selection \cite{mei2024movable}.

\subsection{Pinching-Antenna Systems}
Despite the gains reported for the above-mentioned reconfigurable technologies, the integration of these technologies remains challenging due to the need for real-time control, high implementation cost and complexity, and limited ability to address \textit{large-scale path loss}. To address these challenges, several lines of research have focused on developing alternative efficient reconfigurable technologies for wireless systems. \ac{pas} is one of the most recent advancements in this direction proposed in \cite{ding2024flexible,yang2025pinching}. This proposal is motivated by the demonstration given by Fukuda et al. in DOCOMO 2022 \cite{suzuki2022pinching}, which showcases that a array beam pattern can be constructed by attaching low-cost dielectric materials (referred to as \textit{pinching elements}, such as plastic clothespins, at arbitrary points along a dielectric waveguide. Inspired by this demonstration, \ac{pas} technology suggests to realize large-scale reconfigurable arrays via long dielectric waveguides, in which the propagation pattern is tuned by the location of the pinching elements.

\acp{pas} offer flexibility in establishing or strengthening \ac{los} links by adding or removing pinches to waveguides \cite{ding2024flexible}. Unlike the earlier proposals for reconfigurable arrays, this modification can be easily implemented via significantly low-cost pinching elements. This has attracted several lines of recent research in the literature. The ergodic rate achieved by a \ac{pas} is analyzed \cite{ding2024flexible} using stochastic geometry tools. The array-gain achieved by a \ac{pas} is examined in \cite{ouyang2025array}. Low-complexity algorithms for optimizing the locations of pinching elements along a single waveguide are proposed in \cite{xu2025downlinkpass,wang2024antenna,tegos2024minimum}. These studies collectively highlight the superiority of \acp{pas} over conventional fixed-location antenna systems in enhancing communication performance.






\subsection{Contributions}
The initial study in \cite{ding2024flexible} has investigated the potential of \acp{pas} for \ac{mimo} communication, considering basic multi-user scenarios. The scope of this study is further extended~in recent lines of work, such as \cite{tegos2024minimum,xu2025downlinkpass}. Motivated by these, this work aims to study the potential of \acp{pas} for multiuser \ac{mimo} communication; a gap that has not been explored in the literature. In this respect, we study the problem of multiuser detection and beamforming in the uplink and downlink of a multiuser \ac{mimo} system whose \ac{ap} employs a \ac{pas}-aided transceiver technology. The main lines of contribution in this study are as follows: 
\begin{inparaenum}
\item[($i$)] we consider a generic multi-antenna \ac{pas}, in which the \ac{ap} is equipped with an array of waveguides each being equipped with multiple reconfigurable pinching elements. For this setting, we formulate the uplink and downlink transmission and characterize the effective vector multiple access and broadcast channels between the \ac{pas}-aided \ac{ap} and users.
\item[($ii$)] Using the analytic characterization of the uplink and downlink channels, we develop \textit{hybrid} multi-user detection and beamforming designs, in which the location of the pinching elements on the waveguides and the digital units are designed jointly, such that the weighted achievable sum-rate is maximized.
\item[($iii$)] Invoking \ac{fp}, \ac{bcd} algorithm, and Gauss-Seidel approach, we develop several iterative algorithms for multi-user detection task in uplink and the beamforming task in downlink. The proposed algorithms simultaneously optimize the digital units, i.e., precoding matrix and linear receiver, at the \ac{ap} and adjust the locations of the pinching elements. 
\item[($iv$)] We investigate the efficiency of the proposed designs in both uplink and downlink scenarios through extensive numerical experiments, where we compare different variation of our proposed \ac{pas}-aided transmission with baseline \ac{mimo} schemes; namely, conventional \ac{mimo}, massive \ac{mimo} and hybrid analog-digital architectures.
\end{inparaenum}
Numerical results demonstrate that the proposed \ac{pas}-aided transceiver can achieve significantly higher throughput, as compared with conventional fixed-location antenna systems. This highlights great potentials of this technology to address the targets considered for next generations of wireless systems.  

The remainder of this paper is organized as follows: Section~\ref{sec:formulation} formulates the problem and characterizes the uplink and downlink channels in a \ac{pas}. Hybrid beamforming is studied in Section~\ref{sec:H-DL}, and iterative algorithms are developed in Section~\ref{sec:H-Alg}. Section~\ref{sec:H-UL} develops algorithmic approaches for multiuser detection. Numerical results are presented in Section~\ref{sec:Numerical}. Finally, the paper is concluded in Section~\ref{sec:Conc}.

\subsubsection*{Notation}
Vectors, and matrices are shown by bold lower-case and bold upper-case letters, respectively. The transpose, conjugate and conjugate-transpose of $\mH$ are denoted by $\mH^{\trp}$, $\mH^*$ and $\mH^{\her}$, respectively. The $N\times N$ identity matrix is shown by $\mI_N$, and $\setR$ and $\setC$ denote the real axis and complex plane, respectively. For set $\{1,\ldots,N\}$, we use shortened notation $[N]$.  We denote expectation by $\Ex{.}{}$, and $\mathcal{CN}\brc{\eta,\sigma^2}$ is the complex Gaussian distribution with mean $\eta$ and variance $\sigma^2$. When the summation range is clear from the context, we omit it and denote only the index, e.g., $\sum_{k}$.

\section{System Model}
\label{sec:formulation}
\begin{figure}[!t]
\centering
\includegraphics[height=0.16\textwidth]{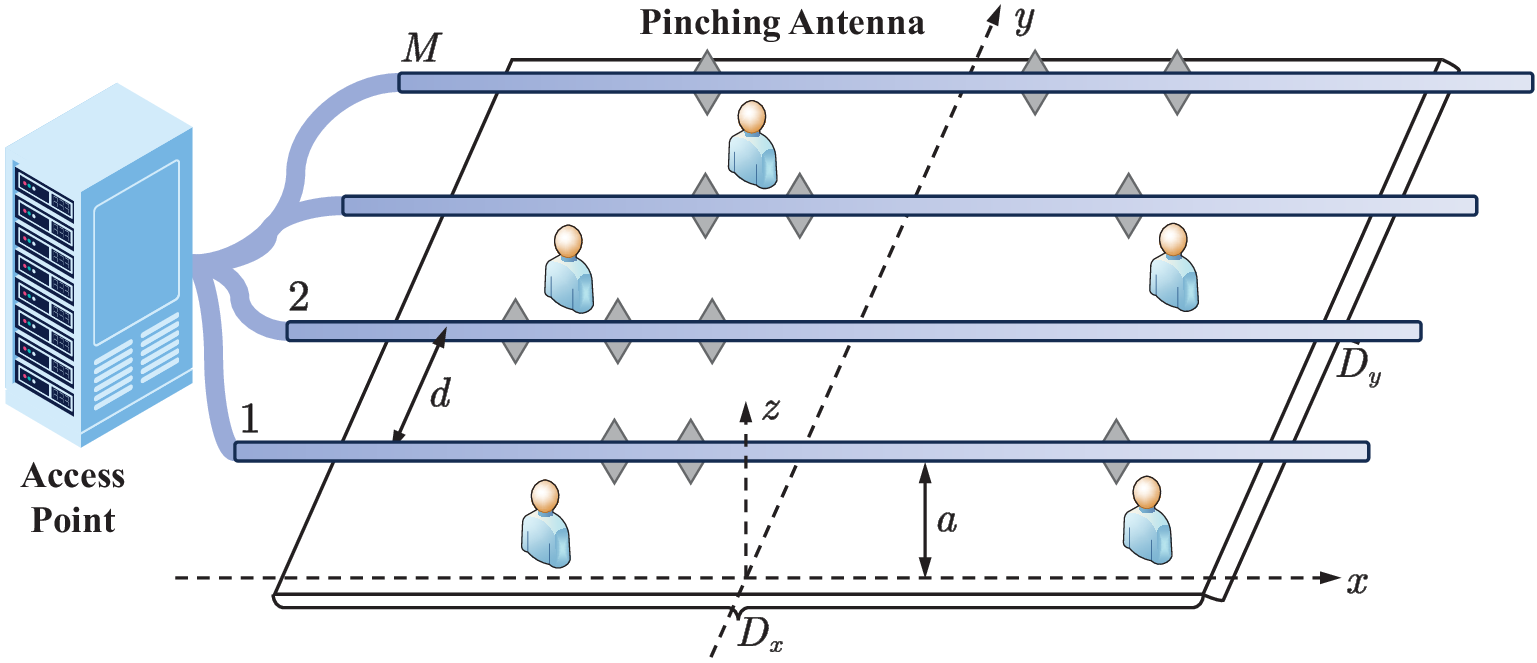}
\caption{Schematic of MIMO-PASS setting: each waveguide is pinched with multiple elements.}\vspace{-5mm}
\label{Fig1_Schematic}
\end{figure}

Consider $M$ dielectric waveguides, each equipped with $N$ pinching elements that can freely move across the waveguide.\footnote{In practice, each element can be realized via multiple elements, each covering one part of the waveguide.} The elements on each waveguide act as an isotropic illuminator whose signals are the phase-shifted version of the signal fed to the waveguide. We assume that waveguides are extended over the $x$-axis at the altitude $a$ in an array formed on the $y$-axis with each two waveguides being distanced $d$. The location of the pinching element $n$ on the waveguide $m\in [M]$ in the Cartesian coordinate is given by $\bv_{m,n} = [\ell_{m,n}, (m-1)d, a]$, where $0 \leq \ell_{m,n} \leq L_m$ is the position of the element on the waveguide $m$ with length $L_m$, assuming that the signal is fed to the wave guide at $\ell_{m,0} = 0$; see Fig.~\ref{Fig1_Schematic}. Note that $\ell_{m,n}$ for $m\in[M]$ and $n\in [N]$ are the design parameters that we tune to optimize the system throughput.

The radiating waveguides are fed by an \ac{ap} that aims to serve $K$ single-antenna users. The users are distributed within a known two-dimensional area, e.g., a rectangular region. We denote the location of user $k$ by $\bu_k = [x_k, y_k, 0]$, assuming that the users are located in the $xy$-plane.

\subsection{Characterizing Downlink Channel}
We start with downlink transmission, in which each waveguide transmits a separate signal to the users. Let $z_m$ denote the signal fed to waveguide $m$. The pinching elements on this waveguide radiate phase-shifted versions of $z_m$. The radiated signal from the pinching elements can hence be represented as $\bt_m = z_m \be_m$, where the \textit{radiation vector} $\be_m\!\in\setC^N$~reads as
\begin{align}
    \be_m = \dbc{p_{m,1}\exp\set{-j \theta_{m,1}}, \ldots, p_{m,N}\exp\set{-j \theta_{m,N}}}^\trp
\end{align}
with $p_{m,n}$ and $\theta_{m,n}$ being the attenuation factor and phase-shift at pinching element $n$ on waveguide $m$. Note that $\theta_{m,n}$ is determined by the location of the element $n$. Recalling that the signal on the waveguide $m$ is fed at $\ell_{m,0} = 0$, and denoting the carrier frequency with $f$, we can write $\theta_{m,n} =  \kappa i_{\mathrm{ref}} {\ell_{m,n}}$,
where $\kappa = 2\pi/\lambda$ is the wave number with $\lambda = \mathrm{c}/f$ denoting the wavelength and $\mathrm{c}$ being the speed of light. The coefficient $i_{\mathrm{ref}}$ is further the refractive index of the waveguides. 

By neglecting the minor propagation loss in the dielectric, each waveguide $m$ can be modeled as a passive power distribution component whose radiated power does not exceed or drop from the one fed to it by $z_m$. This means that we have $\sum_{n} \abs{p_{m,n}}^2 = 1$.
Neglecting the propagation loss further implies that the power radiated from each element is the same. We can hence conclude that in this case $p_{m,n} = \sqrt{1/N}$ for all $n\in[N]$. The radiation vector $\be_m$ is hence simplified as
\begin{align}
    \be_m = \frac{1}{\sqrt{N}} \dbc{\exp\set{-j \theta_{m,1}}, \ldots,\exp\set{-j \theta_{m,N}}}^\trp.
\end{align}

We consider classical indoor scenarios. It is hence practical to assume that users are in the \ac{los} of the waveguides, and that the signals received from the non-\ac{los} paths are negligible in amplitude and separable in delay, as compared with those received over the \ac{los}. Let $y_k$ denote the signal received by user $k$. Considering \ac{los} channels, we can write $y_k$ as 
\begin{align}
    y_k = \sum_{m} \bh_{k,m}^\trp \bt_m + \varepsilon_k
\end{align}
where $\varepsilon_k \in \setC$ is complex-valued additive white Gaussian noise with mean zero and variance $\sigma^2$, i.e., $\varepsilon_k\sim\mathcal{CN} \brc{0,\sigma^2}$, and $\bh_{k,m}\in \setC^N$ denotes the vector of channel coefficients from pinching elements on waveguide $m$ to user $k$. The entry $n$ of the channel vector $\bh_{k,m}$ can be explicitly expressed in terms of the location of the user and the pinching element as
\begin{align} \label{eq:h}
    h_{k,m,n} =\xi \alpha_{k}
    \frac{\exp\set{-j \kappa D_{k,m}\brc{\ell_{m,n}} } }{D_{k,m}\brc{\ell_{m,n}}} 
\end{align}
where $\xi = {\lambda}/{4 \pi}$ is the coefficient proportional to the effective surface of the pinching elements,\footnote{We assume that pinching elements behave as isotropic antennas.} $D_{k,m}\brc{\ell}$ is a function computing the distance between the location $\ell$ on the $m$-th waveguide and user $k$, i.e., 
\begin{align}
D_{k,m}^2\brc{\ell} 
&=(\ell -x_k)^2 + \brc{md - d - y_k}^2 + a^2,
\end{align}
and $\alpha_k$ captures the shadowing experienced by user $k$.

\begin{remark}\label{remark:1}
We assume that shadowing is invariant in $n$. In limited areas, e.g., moderate-size indoor environments, this is a practical approximation \cite{ding2024flexible}. Nevertheless, in a setting where shadowing varies from an element to another, the shadowing coefficients depend on both $m$ and $n$, i.e., $\alpha_{k,m,n}$. 
\end{remark}

The end-to-end channel between the waveguide array and the user $k$ depends on the element locations. Defining location vector of the waveguide $m$ as $\bl_m = [\ell_{m,1}, \ldots, \ell_{m,N}]^\trp$, this means that the channel between waveguide $m$ and user $k$ is a function of $\bl_m$. Considering the radiation vector of waveguide $m$, we can write the end-to-end channel between the waveguide array and the user $k$ as
\begin{align}
    y_k = \sum_{m} g_{k,m} \brc{\bl_m} z_m + \varepsilon_k = {\bg}_{k}^\trp \brc{\mL} \bz + \varepsilon_k
\end{align}
where $\bz = \dbc{z_{1}, \ldots, z_M}^\trp$ denotes the transmitted beam, and $\bg_{k} \brc{\mL} = \dbc{{g}_{k,1} \brc{\bl_1}, \ldots, {g}_{k,M} \brc{\bl_M} }^\trp$ with $\mL = \dbc{\bl_1, \ldots, \bl_M} \in \setR^{N\times M}$ collecting locations of elements and ${g}_{k,m} \brc{\bl_m}\in \setC$ being the effective channel between the waveguide $m$ and the user $k$ given by
\begin{subequations}
\begin{align}\label{eq:g}
   &{g}_{k,m} \brc{\bl_m} = \bh_{k,m}^\trp \be_m\\
    &= \xi \alpha_{k}\sum_{n} 
    \frac{  
    \exp\set{-j \kappa \brc{D_{k,m}\brc{\ell_{m,n}} + i_{\mathrm{ref}} \ell_{m,n} } }
    }{\sqrt{N} D_{k,m}\brc{\ell_{m,n}} } .
\end{align}
\end{subequations}

The vector downlink channel is hence given by 
\begin{align}
    \by = \mG^\trp \brc{\mL} \bz + \bepsilon,
\end{align}
where $\mG\!\brc{\mL}\!\in\!\setC^{M\!\times\!K}$ is the downlink channel defined as
\begin{align}\label{eq:G_mat}
    \mG\brc{\mL} = \dbc{\bg_{1} \brc{\mL} \; \cdots \; \bg_{K} \brc{\mL}  },
\end{align}
and $\bepsilon\in\setC^K$ is Gaussian noise. This describes a standard vector Gaussian broadcast channel whose channel matrix is tunable by the activated locations of the pinching elements.

\subsection{Characterizing Uplink Channel}
\label{sec:CH-ul}
We now consider uplink transmission in which the array of waveguides receives the user signals. Let us denote the uplink signal transmitted by user $k$ via $x_k$. Assuming that the system operates in the \ac{tdd}, the received signal by pinching element $n$ on waveguide $m$ is given by
\begin{align}
    r^{\rm e}_{m,n} = \sum_{k} h_{k,m,n} x_k + \nu_{m,n}^{\rm e}
\end{align}
where $h_{k,m,n}$ is the channel coefficient between user $k$ and pinching element $n$ on waveguide $m$ given in \eqref{eq:h}, and $\nu_{m,n}^{\rm e}$ is Gaussian noise with variance $\varsigma^2$, i.e., $\nu_{m,n}^{\rm e} \sim \mathcal{CN} \brc{0, \varsigma^2}$.

Following our assumption on passiveness of the waveguides, we can write the superimposed signal received through waveguide $m$ at the \ac{ap} as
\begin{subequations}
\begin{align}
    \tilde{r}_{m} &= \sum_{n} \exp\set{-j \kappa \ell_{m,n}}r^{\rm e}_{m,n}\\
    &= \sum_{k} \sum_{n}  h_{k,m,n} \exp\set{-j \kappa \ell_{m,n}} x_k + \tilde{\nu}_{m},
\end{align}
\end{subequations}
where $\tilde{\nu}_{m}$ is aggregated noise process, i.e., $ \tilde{\nu}_{m} = \sum_{n} \nu_{m,n}^{\rm e}$.

The independence of $\nu_{m,n}^{\rm e}$ implies that $\tilde{\nu}_{m}$ is zero-mean Gaussian with variance $N \varsigma^2$. Considering \eqref{eq:g}, we can write
\begin{align}
    \sum_{n}  h_{k,m,n} \exp\set{-j \kappa \ell_{m,n}} = \sqrt{N} g_{k,m} \brc{\bl_m}.
\end{align}
Hence, the signal received by the \ac{ap} can be written as 
\begin{align}
    \hspace{-.2cm}\tilde{r}_{m}\!=\!\sqrt{N} \sum_{k=1}^K g_{k,m} \brc{\bl_m} x_k\!+\!\tilde{\nu}_{m}\!=\!\sqrt{N} \tilde{\bg}_m^\trp \brc{\bl_m} \bx + \tilde{\nu}_m,
\end{align}
where $\tilde{\bg}_m \brc{\bl_m} = [g_{1,m} \brc{\bl_m} \; \cdots \; g_{K,m} \brc{\bl_m}]^\trp$ is the $m$-th row of $\mG \brc{\mL}$ defined in \eqref{eq:G_mat}, and $\bx = \dbc{x_1 \; \cdots \; x_K}$ represents the vector of uplink signals. Noting that scaling at the received side does not impact transmission quality, we can  compactly represent the effective uplink channel as
\begin{align}
    \br
    &= \mG \brc{\mL} \bx + \bnu,
\end{align}
where $\br = \dbc{r_1 \; \cdots \; r_M}^\trp$ with $r_m = {\tilde{r}_{m}}/{\sqrt{N}}$ representing the effective received signal, and $\bnu = \dbc{\nu_1 \; \cdots \; \nu_M}^\trp$ with $\nu_m = \tilde{\nu}_m/\sqrt{N}$ being effective zero-mean noise with variance $\varsigma^2$. 

As a dual case to downlink transmission, the end-to-end uplink channel describes a Gaussian multiple access channel. The channel matrix in this case is the transpose of the downlink channel, which is the direct result of channel \textit{reciprocity} in the \ac{tdd} mode. Similar to downlink case, the key advantage in this system is the reconfigurability of the channel matrix by the locations of the pinching elements. 

\section{Hybrid Beamforming in a Downlink PASS}
\label{sec:H-DL}
Considering the equivalent end-to-end channel for the \ac{pas} system, we can achieve multiplexing gain via digital precoding. The design of precoding however describes a \textit{hybrid} beamforming problem, in which both the \textit{digital precoder} and \textit{locations of the pinching elements} are to be optimized jointly. 

\subsection{Joint Digital Precoding and Location Tuning}
Let $\bs = [s_1, \ldots,s_K]^\trp$ denote the signals the \ac{ap} intends to send, with $s_k\in \setC$ being the encoded information signal of user $k$. We assume that the information signals are zero-mean unit-variance stationary processes satisfying
\begin{align}
    \Ex{\bs\bs^\her} = \mI. \label{eq:iid_s}
\end{align}
The \ac{ap} sets the signal fed to each waveguide to be a linear superposition of these signals, i.e., it sets
    $z_m = \Tilde{\bw}_m^\trp \bs$, %
for some vector of coefficients $\Tilde{\bw}_m \in \setC^K$ specified for waveguide $m$. The transmit signal by the waveguide array can hence be represented as
    $\bz = \mW \bs$, 
where $\mW = \dbc{\Tilde{\bw}_1, \ldots, \Tilde{\bw}_M}^\trp$ is the precoding matrix. We further impose a power constraint restricting the average transmit power to be bounded by a positive downlink power $P_{\rm d}$, i.e., $ \Ex{\norm{\bz}^2} \leq P_{\rm d}$. Using \eqref{eq:iid_s}, this reduces to $\tr{\mW \mW^\her} \leq P_{\rm d}$.

The signal received by user $k$ is written as
\begin{align}
    y_k &= \bg_{k}^\trp \brc{\mL} \mW \bs + \varepsilon_k 
\end{align}
which depends on $\mW$ and $\mL$. It is worth mentioning that while $\mW$ represents a \textit{digital} beamforming, the tuning of matrix $\mL$ describes beam-design in the \textit{analog} domain. The joint design of $\mW$ and $\mL$ can hence be interpreted as hybrid beamforming.

\subsection{Downlink Weighted Sum-Rate}
To proceed with the beamforming design, let us first rewrite the received signal at user $k$ as
\begin{align}
    y_k &= \sum_{j} \bg_{k}^\trp\brc{\mL} \bw_j s_j + \varepsilon_k
\end{align}
with $\bw_j$ denoting the $j$-th column vector of $\mW$ collecting the $j$-th entries of $\Tilde{\bw}_m$ for $m\in [M]$. We refer to $\bw_j$ as the \textit{digital beamforming vector} of user $j$. The achievable downlink rate for user $k$ in this case can be written as
\begin{align}
    R_k^\dl \brc{\mW, \mL} = \log \brc{1 + \sinr_k^\dl \brc{\mW, \mL}},
\end{align}
where $\sinr_k^\dl\brc{\mW, \mL}$ denotes the \ac{sinr} at user $k$ defined as
\begin{align}
    \sinr_k^\dl \brc{\mW, \mL} = \frac{\abs{\bg_k^\trp  \brc{\mL} \bw_k}^2}{\displaystyle\sum_{j\neq k} \abs{\bg_k^\trp \brc{\mL} \bw_j}^2 + \sigma^2 }.
    \label{eq:SINR_k}
\end{align}
The achievable weighted sum-rate in the downlink vector channel can hence be written as
\begin{align}
    \R^\dl \brc{\mW, \mL} = \sum_{k} \lambda_k R_k^\dl \brc{\mW, \mL},
\end{align}
for some non-negative weights $\lambda_k$ that are proportional to the expected quality of service for user $k$. 

\subsection{Optimal Beamforming Design}
The ultimate goal is to maximize the achievable weighted sum-rate for a given power budget $P_\dl$ and physical limitations. We hence formulate optimal hybrid beamforming as 
\begin{align}
    &\max_{\mW, \mL} \R^\dl \brc{\mW, \mL}  \label{eq:optim-old}
    \subto \tr{\mW^\her \mW} \leq P_\dl \txtand \mathcal{C}_1 , \mathcal{C}_2 
\end{align}
where we define $\mathcal{C}_1$ and $\mathcal{C}_2$ as
\begin{subequations}
\begin{align}
&\mathcal{C}_1: 0 \leq \ell_{m,n}\leq L_m\\
&\mathcal{C}_2: \Delta\ell + \ell_{m,n} \leq \ell_{m,n+1}, \label{C2-old}
\end{align}
\end{subequations}
for $n\in[N]$ and $m\in[M]$ recalling that $L_m$ denotes the length of the waveguide $m$. 

In addition to restricting the location of pinching elements with the length of waveguides, i.e., $\mathcal{C}_1$, $\mathcal{C}_2$ guarantees that the element $n$ is located behind element $n+1$ with a distance that exceeds $\Delta\ell$. This is physically the case when pinching elements are tuned by sliding tracks on waveguides. The parameter $\Delta\ell>0$ is the minimum spacing required to prevent mutual coupling between two neighboring elements.

Although the formulation in \eqref{eq:optim-old} matches the physical~limitations of the \ac{pas}, one can observe that the second constraint in \eqref{C2-old}, i.e., $\Delta\ell + \ell_{m,n} \leq \ell_{m,n+1}$, can be further simplified. This is intuitive, as exchanging two pinching elements on a waveguide in this case does not change signal propagation.~In other words, the throughput is invariant to permutation of pinching elements. This is shown in the following lemma.
\begin{lemma}\label{lem:perm}
    Consider $\mL \in \setR^{N\times M}$. Let $\tilde{\mL}\in \setR^{N\times M}$ be another location matrix whose columns are a permutation of the columns of $\mL$. Then, for any $\mW \in \setC^{M\times K}$ we have
    \begin{align}
    \R^\dl \brc{\mW, \mL} = \R^\dl (\mW, \tilde{\mL}).
\end{align}
\end{lemma}
\begin{IEEEproof}
    From \eqref{eq:g}, we see that by replacing $\bl_m$ with a permutation $\tilde{\bl}_m$, the sum does not change implying $g_{k,m} \brc{\bl_m} = g_{k,m} (\tilde{\bl}_m)$. Noting that $\R^\dl \brc{\mW, \mL}$ depends on the locations only through the channels, the proof is completed.
\end{IEEEproof}

Using this permutation invariance, the optimal hybrid precoding scheme simplifies to
\begin{subequations}
\begin{align}
    &\max_{\mW, \mL} \R^\dl \brc{\mW, \mL}  \label{eq:optim}\\
    &\subto \tr{\mW^\her \mW} \leq P_\dl \txtand  \bl_m \in \setL_m, 
\end{align}
\end{subequations}
where $\bl_m$ denotes the $m$-th column of $\mL$, and we define the feasible set $\setL_m$ for $m\in\dbc{M}$ as
\begin{align}\label{setL}
    \setL_m = &\left\lbrace \bl \in \setR^N: 0 \leq \ell_n \leq L_m  \txtand \nonumber \right. \\
    &\qquad \left. \qquad\abs{\ell_n - \ell_{n'}} \geq \Delta \ell \txtfor n,n'\in\dbc{N} \right\rbrace
\end{align}
with $\ell_n$ in \eqref{setL} denoting the $n$-th entry of $\bl$. 
\begin{remark}
It is worth mentioning that the permutation invariance is implied by the invariance of the shadowing across the pinching elements. In fact, when the shadowing varies, i.e., we have path-loss coefficients of the form $\alpha_{k,m,n}$ that vary with $n$, one needs to solve the original problem in \eqref{eq:optim-old} for optimal hybrid beamforming. See Remark~\ref{remark:1} for more details.
\end{remark}

\section{Efficient Algorithm for Hybrid Beamforming}
\label{sec:H-Alg}
The optimization in \eqref{eq:optim} is non-convex, and its global solution is not computed in polynomial time. In this section, we develop an efficient suboptimal algorithm. The derivation follows three major steps: ($i$) we first rewrite \eqref{eq:optim} in a variational form with no power constraint. ($ii$) Using \ac{fp}, we transform the variational problem into a dual optimization with quadratic objective. ($iii$) Using \ac{bcd} \cite{tseng2001convergence} and Gauss-Seidel \cite{sauer2011numerical} methods, a two-tier iterative solver is developed. 

\subsection{Variational Design Problem}
We start with transforming the problem \eqref{eq:optim} into a \textit{variational} form without the power constraint, whose dual form exhibits a smoother objective function. The variational optimization is derived using the following lemmas.
\begin{lemma}[Equality power constraint]\label{Equality Power Constraint}
The optimal solution to\eqref{eq:optim} satisfies the power constraint with equality, i.e., 
\begin{align}\label{equality power constraint}
\tr{\mW_\star^{\her} \mW_\star} = P_\dl,
\end{align}
for any $\mW_\star$ that is a solution to \eqref{eq:optim}.
\end{lemma}
\begin{IEEEproof}
The proof is concluded by contradiction: let the precoding matrix $\hat{\mathbf{W}}=[{\hat{\mathbf{w}}}_1,\ldots,{\hat{\mathbf{w}}}_K]$ be a solution to problem \eqref{eq:optim} that satisfies the constraint with inequality, i.e., 
\begin{align}
\hat{P}\triangleq \tr{\hat{\mathbf{W}}^\her\hat{\mathbf{W}}}=\sum_{j}\norm{{\hat{\mathbf{w}}}_j}^2<P_\dl.    
\end{align}
We now define $\rho = P_\dl/\hat{P}$ and a scaled solution $\bar{{\mathbf{w}}}_j=\sqrt{\rho}{\hat{\mathbf{w}}}_j$. Note that the scaled matrix $\bar{\mW}= [{\bar{\mathbf{w}}}_1,\ldots,{\bar{\mathbf{w}}}_K]$ satisfies the power constraint with equality, i.e., $\tr{\bar{\mW}^{\her} \bar{\mW}} = P_\dl$. Replacing into the definition \eqref{eq:SINR_k}, it is readily shown that 
\begin{align}
    \sinr_k^\dl \brc{\bar{\mW}, \mL} > \sinr_k^\dl (\hat{\mW}, \mL),
\end{align}
for any $\mL$. This means that using $\bar{\mW}$, the \ac{ap} can achieve a strictly larger weighted sum-rate. This contradicts the initial assumption 
and completes the proof.
\end{IEEEproof}

Using Lemma~\ref{Equality Power Constraint}, we can rewrite the optimization in \eqref{eq:optim} in an unconstrained form whose solution determines explicitly the solution of \eqref{eq:optim}. This is shown in the following lemma.

\begin{lemma}[Unconstrained Equivalent Problem]\label{lem:3}
For a given $\mL$, let $\bar{\mW}= [{\bar{\mathbf{w}}}_1,\ldots,{\bar{\mathbf{w}}}_K]$ denote a solution to the following unconstrained weighted sum-rate maximization
\begin{align}\label{Beamforming_Problem_2}
\max_{\mW}  \sum_{k} \lambda_k \log(1+\overline{\sinr}_k^\dl \brc{\mW,\mL}),
\end{align}
where $\overline{\sinr}_k^\dl \brc{\mW,\mL}$ for ${\mW}= [{{\mathbf{w}}}_1,\ldots,{{\mathbf{w}}}_K]$ is defined as
\begin{align}\label{eq:sinr_k_bar}
\overline{\rm{SINR}}_k^\dl \brc{\mW,\mL} =\frac{\abs{\bg_k^\trp\brc{\mL}  \bw_k}^2}{\displaystyle \sum_{j\neq k} \abs{\bg_k^\trp\brc{\mL}  \bw_j}^2 + \frac{\sigma^2}{P_\dl}\sum_{j}\norm{{{\mathbf{w}}}_j}^2 }.
\end{align}
Then, a solution to \eqref{eq:optim} is determined from $\bar{\mW}$ as
\begin{align}\label{Scaled_Solution}
\breve{\mW} =\sqrt{\frac{P_\dl}{\tr{\bar{\mW}^{\her} \bar{\mW}}}} \bar{\mW}.
\end{align}
\end{lemma}
\begin{IEEEproof}
It is readily shown that the solution in \eqref{Scaled_Solution} satisfies the power constraint with equality and that 
\begin{align}
\overline{\rm{SINR}}_k^\dl \brc{\breve{\mW},\mL} = {\rm{SINR}}_k^\dl \brc{\breve{\mW},\mL},
\end{align}
for any $\mL$. This implies that the objective function in \eqref{eq:optim} attains the same value as the one in \eqref{Beamforming_Problem_2}. Noting that $\bar{\mW}$ is the optimal solution to \eqref{Beamforming_Problem_2}, we can conclude that $\breve{\mW}$ maximizes the objective in \eqref{eq:optim} among all beamforming matrices that satisfy the power constraint with equality. Lemma \ref{Equality Power Constraint} further indicates that no solution with inequality constraint exists. This implies that $\breve{\mW}$ maximizes the objective in \eqref{eq:optim}. This completes the proof. 
\end{IEEEproof}

Using Lemma~\ref{lem:3}, the optimal hybrid beamforming is equivalently represented by the following variational optimization 
\begin{subequations}
\begin{align}\label{Beamforming_Problem_simple}
    &\max_{\mW, \bl} \bar{\R}^\dl \brc{\mW, \mL}
    \subto \bl_m \in \setL_m 
\end{align}
\end{subequations}
where $\bar{\R}^\dl \brc{\mW, \mL}$ is given by
\begin{align}
\bar{\R}^\dl \brc{\mW, \bl}&=\sum_{j} \lambda_k \log \brc{1+\overline{\sinr}_k^\dl \brc{\mW, \mL}}
\end{align}
with $\overline{\sinr}_k^\dl \brc{\mW, \mL}$ being defined in \eqref{eq:sinr_k_bar}. It is worth noting that the power limit in the variational form is incorporated through $\overline{\sinr}_k^\dl \brc{\mW, \mL}$ and the optimal precoding matrix is computed from the solution by the scaling given in \eqref{Scaled_Solution}.

\subsection{Solution via Fractional Programming}
The variational problem in \eqref{Beamforming_Problem_simple} describes a classical problem of maximizing sum of log ratios, whose solution can be efficiently approximated via \ac{fp} as outlined in this sub-section. We start our derivations by defining the feasible set $\setS$ for the variational problem as
\begin{align}\label{eq:setS}
\setS = \setC^{M\times K} \times \setL_1 \times \cdots \times \setL_M,
\end{align}
and denoting $\mS = \brc{\mW, \mL}$ for brevity. 

The \ac{fp} approach suggests converting the sum-of-log-ratios problem, i.e., the variational problem \eqref{Beamforming_Problem_simple}, into a quadratic form by means of the Lagrange and quadratic dual transforms \cite{shen2018fractional1,shen2018fractional2}. Although the resulting dual problem is yet non-convex, it describes a smooth objective landscape, whose optimal point can be efficiently approximated by \ac{bcd} algorithm \cite{tseng2001convergence}. For a quick introduction to \ac{fp}, see \cite[Section IV]{asaad2022secure}. In the sequel, we follow this approach: we start with determining the Lagrange dual objective for \eqref{Beamforming_Problem_simple}, which is given by \eqref{eq:dual_dl_1} given at the top of the next page
\begin{figure*}
\begin{align}
\Lag^\dl \brc{\mS, \bomega} = \sum_{k} \lambda_k 
\brc{
\log\brc{1+\omega_k} - \omega_k + 
\frac{\brc{1+\omega_k} \abs{\bg_k^\trp \brc{\mL} \bw_k}^2 }{\displaystyle \sum_j \abs{\bg_k^\trp\brc{\mL} \bw_j}^2 
+ \frac{\sigma^2}{P}\sum_{j}\norm{{{\mathbf{w}}}_j}^2
}
}\label{eq:dual_dl_1}
\end{align}
\hrule
\end{figure*}
for some $\omega_n \geq 0$. The key property of this dual objective is the Lagrange duality, which implies that $\mS_{\max}$ for the dual problem
\begin{align}
    \brc{\mS_{\max}, \bomega_{\max}} = \argmax_{\mS \in \setS, \bomega \in \setR_+^K} \Lag^\dl \brc{\mS, \bomega}
\end{align}
recovers the solution of the optimization \eqref{Beamforming_Problem_simple}, and that the optimal values of the objectives in both problems are identical \cite{shen2018fractional1,shen2018fractional2}. Note that the Lagrange dual problem is not convex. Nonetheless, its solution can be efficiently approximated via the \ac{bcd} algorithm: we initiate some $\mS = \brc{\mW, \mL}$ and iterate between the following two marginal optimization problems.
\begin{enumerate}
    \item For fixed $\mS$, solve the marginal optimization
    \begin{align}
     \bomega^\star = \argmax_{\bomega \in \setR_+^K} \Lag^\dl \brc{\mS, \bomega}.
\end{align}
This problem describes a standard convex optimization whose solution is given by 
\begin{align}\label{fp_bcd_omage_k}
  \omega_k^\star = \overline{\sinr}_k^\dl \brc{\mW, \mL}.  
\end{align}
\item Set $\omega_k = \omega_k^\star$, i.e., the solution of first marginal problem, and update $\mS$ with the marginal solution $\mS^\star$ given by
 \begin{align}
     \mS^\star &= \argmax_{\mS \in \setS} \Lag^\dl \brc{\mS, \bomega^\star}
\end{align}
\end{enumerate}
The second marginal optimization reduces to
 \begin{align}
     \mS^\star 
     &= \argmax_{\mS \in \setS} 
\sum_{k}
\frac{\lambda_k \brc{1+\omega_k^\star} \abs{\bg_k^\trp \brc{\mL} \bw_k}^2 }{ \displaystyle \sum_j \abs{\bg_k^\trp\brc{\mL} \bw_j}^2 
+ \frac{\sigma^2}{P_\dl}\sum_{j}\norm{{{\mathbf{w}}}_j}^2}, \label{eq:Fractional}
\end{align}
which describes the problem of maximizing sum of fractional functions. This can be transformed into a quadratic form using the \textit{quadratic transform} \cite{shen2018fractional1,shen2018fractional2}, as outlined in the sequel.

To solve the optimization in \eqref{eq:Fractional}, we determine its quadratic dual objective, which is given by \eqref{eq:Quadratic} at the top of the next page for some complex auxiliary variables $q_k \in \setC$ \cite{shen2018fractional1,shen2018fractional2}.
\begin{figure*}[t]
\begin{align}
     \Q^\dl \brc{\mS, \bq} = 
\sum_{k} \lambda_k
\brc{2 \sqrt{1+\omega_k^\star}\Re \set{ q^*_k \bg_k^\trp \brc{\mL} \bw_k } - \abs{q_k}^2 \sum_{j} \abs{\bg_k^\trp\brc{\mL} \bw_j}^2 -  \frac{\sigma^2}{P_\dl} \abs{q_k}^2
\sum_{j}\norm{{{\mathbf{w}}}_j}^2
}\label{eq:Quadratic}
\end{align}
\hrule
\end{figure*}
The quadratic duality implies that the solution of \eqref{eq:Fractional} is recovered by solving the dual problem \cite{shen2018fractional1,shen2018fractional2}
 \begin{align}
     \max_{\mS \in \setS, \bq \in \setC^K} 
     \Q^\dl \brc{\mS, \bq}.
\end{align}
This dual form has a quadratic objective which describes a smoother landscape, and hence its solution can be approximated more efficiently. Note that the dual optimization is again non-convex through $\bq$, multiplicative expressions, and the functional form of $\bg_k\brc{\mL}$. Similar to the Lagrange dual problem, we approximate the dual solution via \ac{bcd}: starting with $\mS$, the algorithm iterates between the following two steps.
\begin{enumerate}
    \item Fix $\mS$ and find $\bq^\star$ by solving \eqref{eq:Quadratic} marginally for $\bq$. 
    \item Update $\mS$ by solving \eqref{eq:Quadratic} marginally for $\mS$, i.e., 
        \begin{align}
        \mS^\star = \argmax_{\mS \in \setS} 
     \Q^\dl \brc{\mS, \bq^\star}
    \end{align}
\end{enumerate}
The solution to the first marginal problem is given by
    \begin{align}\label{fp_bcd_q_k}
        q_k^{\star} = \frac{ P_\dl \sqrt{1+\omega_k^\star}
        \bg_k^\trp \brc{\mL}  \bw_k 
        }{
        \sigma^2 \tr{\mW \mW^\her} + P_\dl  \tr{\mW \mW^\her \bg_k^*\brc{\mL}  \bg_k^\trp \brc{\mL} }
        }.
    \end{align}
The second marginal problem can be further written as
    \begin{align}\label{eq:lW_problem_Final}
        \mS^\star = \argmax_{\brc{\mW,\mL} \in\setS} 
         F_\dl \brc{\mW, \mL},
    \end{align}
    where the objective $F_\dl \brc{\mW, \mL}$ is given by
    \begin{align}
        &F_\dl\brc{\mW, \mL} = 2 \Re\set{ \tr{\mT^\her \mG^\trp \brc{\mL} \mW} } \\
        &- \tr{ \mG^\trp \brc{\mL} \mW\mW^\her \mG^* \brc{\mL} \mU  } 
        - \frac{\sigma^2\tr{\mU}}{P_\dl}\tr{\mW\mW^\her}\nonumber
    \end{align}
    for $\mT = \mQ\mA \mLambda $ and $\mU = \mQ \mLambda \mQ^\her$ with $\mLambda$, $\mA$ and $\mQ$ being
    \begin{subequations}\label{eq:U-T}
    \begin{align}
        \mLambda &= \diag{\lambda_1, \ldots, \lambda_K}\\
        \mA &= \diag{\sqrt{1+\omega_1^\star}, \ldots, \sqrt{1+\omega_K^\star}}\\
        \mQ &= \diag{q_1^\star, \ldots, q_K^\star}.
    \end{align}
    \end{subequations}

The final step is to approximate the solution of the marginal problem in \eqref{eq:lW_problem_Final}. This is challenging due to two facts: ($i$) it is non-convex, and ($ii$) it shows too many local minima, due to its oscillating form. We address these challenges by developing an iterative solver via the \ac{bcd} and Gauss-Seidel schemes.

\subsection{Two-tier Iterative Algorithm}\label{sec:2tier-dl}
To solve the optimization in \eqref{eq:lW_problem_Final}, we first note that the problem is marginally convex in both $\mW$. This suggests that a \ac{bcd}-based outer loop can efficiently approximate the optimal digital precoding matrix: the outer loop iterates between the following two steps.
\begin{enumerate}
    \item Fix $\mL$, and find $\mW^\star$ as the marginal solution to \eqref{eq:lW_problem_Final}. 
    
    \item Set $\mW=\mW^\star$ and solve \eqref{eq:lW_problem_Final} marginally for $\mL$. 
    \end{enumerate}

    The first marginal problem describes a standard regularized linear inverse problem whose solution is given by 
            \begin{align}
        {\mathbf{W}}^{\star}
        &=\brc{\mG^* \brc{\mL} \mU \mG^\trp \brc{\mL}+\gamma_\dl \mI_M}^{-1}\mG^* \brc{\mL}\mT.\label{fp_bcd_W}
    \end{align}
    with $\gamma_\dl = {\sigma^2\tr{\mU}}/{P_\dl}$.
    \begin{remark}
        The digital precoder \eqref{fp_bcd_W} can be seen as \ac{rzf} precoding with an effective channel given in terms of $\mG\brc{\mL}$, $\mT$, and $\mU$ whose regularizer is set to ${\sigma^2\tr{\mU}}/{P_\dl}$. This is intuitive, as for a fixed channel, i.e., fixed location $\mL$, the optimal linear precoder is readily computed by an optimally-tuned \ac{rzf}. 
    \end{remark}

The second marginal problem in the outer loop deals with an objective whose dependency on $\mL$ is given in terms of exponential sums. The global optimum of this problem is not feasibly found. We hence develop an inner-loop algorithm to iteratively approximate its solution. The classical approach to tackle these problems is to invoke greedy approaches, e.g., stepwise regression. In the sequel, we develop a greedy approach based on the Gauss-Seidel scheme to efficiently approximate the solution of this marginal problem. 

\subsection{Greedy Algorithm for Location Optimization}
\label{sec:GaussSeidel}
The inner loop solves the following optimization 
    \begin{subequations}\label{loc_Optim1}
    \begin{align}\label{loc_Optim}
        &\max_{\mL} 2 \Re\set{ \tr{\mE \mG^\trp \brc{\mL} } } - \tr{ \mF \mG^* \brc{\mL} \mU \mG^\trp\brc{\mL}}\\
        &\subto {\bl_m\in \setL_m \txtfor m\in [M] },
    \end{align}
    \end{subequations}
where we define $\mE = \mW \mT^\her$ and $\mF = \mW\mW^\her$.

To tackle the problem in \eqref{loc_Optim1}, we invoke the Gauss-Seidel approach, which suggests to sequentially update the locations \cite{sauer2011numerical}. In this approach, each location is updated individually by treating other locations as constants. This way, each location is optimized to \textit{refine} analog beamforming optimized in previous iterations. To this end, let us consider the element $n$ on waveguide $m$ and fix $\ell_{m',n'}$ for ${(m',n')\ne (m,n)}$. The scalar problem for this element is given by
    \begin{align}
        &\max_{ \ell } 2 \Re\set{ \tr{\mE \tilde{\mG}_{m,n}^\trp \brc{\ell} } } - \tr{ \mF \tilde{\mG}_{m,n}^* \brc{\ell} \mU \tilde{\mG}_{m,n}^\trp \brc{\ell} } \nonumber \\
        &\subto  0 \leq \ell \leq L_m \txtand 
        \Delta\ell \leq \lvert\ell_{m,i}-\ell \rvert 
        \txtfor i\ne n,\label{Optimization_single_location}
    \end{align}
    where $\tilde{\mG}_{m,n} \brc{\ell}$ represents matrix $\mG\brc{\mL}$, when all locations but $\ell_{m,n} = \ell$ are set to their fixed values, i.e., 
    \begin{align}\label{eq:G_tild}
    \tilde{\mG}_{m,n} \brc{\ell}=[\tilde{\bg}_1,\ldots,\tilde{\bg}_{m-1},\tilde{\bg}_m^n\brc{\ell},\tilde{\bg}_{m+1},\ldots,\tilde{\bg}_{M}]^\trp,
    \end{align}
    where $\tilde{\bg}_{m'}$ for $m'\neq m$ represents the $m'$-th rows of $\mG\brc{\mL}$, when all entries of $\mL$ except $\ell_{m,n}$ are set to their fixed values, and $\tilde{\bg}_{m}^n \brc{\ell}$ is the $m$-th row as a function of $\ell_{m,n} = \ell$.\footnote{Noting that $\ell_{m',n}$ for ${m'\ne m}$ is treated as fixed, the argument $\ell_{m',n}$ in $\tilde{\bg}_{m'}$ for ${m'\ne m}$ is dropped. We also use superscript $n$ in $\tilde{\bg}_{m}^n \brc{\ell}$ to indicate that that $\ell$ is the location of the $n$-th element.} Note that the scalar objective of \eqref{Optimization_single_location} depends on the optimization variable only through $\tilde{\bg}_{m} \brc{\ell}$ whose $k$-th entry is given by
    \begin{align}\label{definition_h_l_m}
    \dbc{\tilde{\bg}_m^n\brc{\ell}}_k 
    &=
    \Pi_{k,m} \brc{\ell} + \sum_{n'\neq n} \Pi_{k,m} \brc{\ell_{m,n'}}
    \end{align}
    where we define the function $\Pi_{k,m}\brc{\ell}$ as
    \begin{align}
    \Pi_{k,m}\brc{\ell} &= \xi \alpha_{k} 
    \frac{  
    \exp\set{-j \kappa \brc{D_{k,m}\brc{\ell} + i_{\mathrm{ref}} \ell } }
    }{\sqrt{N} D_{k,m}\brc{\ell } }.
    \end{align}

   By simple lines of derivation, \eqref{Optimization_single_location} is rewritten as 
    \begin{align}\label{Optimization_single_location_simplified}
        &\max_{ \ell}  \;
        2\Re\set{ \bb_m^{\mathsf{T}}\tilde{\bg}_{m}^n\brc{\ell} } 
        - [\mF]_{m,m} \tilde{\bg}_{m}^{n\her}\brc{\ell}\mU \tilde{\bg}_{m}^n \brc{\ell} \\ 
         &\subto  0 \leq \ell\leq L_m \txtand 
        \Delta\ell \leq \lvert\ell_{m,i}-\ell\rvert 
        \txtfor i\ne n,
    \end{align}
    where $\bb_m \in \setC^K$ is defined as
    \begin{align}\label{eq:bbm}
    \bb_m = \ba_m - \sum_{m'\ne m}[\mF]_{m,m'} \mU^\trp \tilde{\bg}_{m'}^{*},
    \end{align}
    with $\ba_m^{\trp}$ being the $m$-th row of $\mE$. Using \eqref{definition_h_l_m} and treating  $\ell_{m',n'}$ for $\brc{m',n'} \neq \brc{m,n}$ as fixed, we can simplify \eqref{Optimization_single_location_simplified} to
    \begin{subequations}
    \begin{align}\label{Optimization_single_location_transformed}
        &\max_{ \ell}  \;
        f_{m}(\ell )  \\ 
         &\subto  0 \leq \ell \leq L_m \txtand 
        \Delta\ell \leq \lvert\ell_{m,i}-\ell \rvert 
        \txtfor i\ne n,
    \end{align}
    \end{subequations}
    where $f_{m}(\ell )$ is defined as
    \begin{align}\label{eq:fm}
    f_{m}(\ell ) 
     &=
        \sum_{k}
    2 \Re \set{ \zeta_{k,m} \Pi_{k,m} \brc{\ell } } 
    -\vartheta_{k,m}  \abs{\Pi_{k,m} \brc{\ell }}^2.
    \end{align}
for scalars $\zeta_{k,m} = \dbc{\bb_m}_k$ and $\vartheta_{k,m} = [\mF]_{m,m} [\mU]_{k,k}$. This is a classical scalar optimization problem within a fixed interval, which can be effectively solved via grid search. Using an $L$-point grid for interval $[0,L_m]$, i.e.,
    \begin{align*}
    \hat{\setL}_m \triangleq \set{0,\frac{L_m}{L-1},\frac{2L_m}{L-1},\ldots,L_m}    
    \end{align*}
    a nearly-optimal $\ell_{m,n}$ is computed as
    \begin{align}\label{fp_bcd_l_m}
    \ell_{m,n}^\star =\argmax_{\ell_m\in \hat{\setL}_m\setminus \breve{\setL}_{m,n}} f_{m}(\ell_{m,n}),
    \end{align}
    where the set $\breve{\setL}_{m,n}$ includes all grid points that violate the minimum distance criteria, i.e., 
    \begin{align}
    \breve{\setL}_{m,n} = \hat{\setL}_m\cap &\left\lbrace \ell : 0 \leq \ell\leq L_m \txtand \right. \\ &\left. \Delta\ell \geq \lvert\ell_{m,i}-\ell \rvert \txtfor i\ne n \right\rbrace . \nonumber
    \end{align}
Consequently, the optimal locations in $\mL$ are computed by looping this grid reach over all entries of $\mL$. 


\begin{remark}
Note that classical gradient-based methods cannot solving the scalar problem in \eqref{Optimization_single_location_transformed} efficiently. This is due to the fact taht the objective $f_{m}(\cdot)$ contains numerous stationary points caused by oscillations of the cosine term.
\end{remark}

\subsection{Final Algorithm: Convergence and Complexity}
Using the Gauss-Seidel approach, we can complete the inner-loop of the two-tier solver: in Step 2 of the \ac{bcd} loop in Section~\ref{sec:2tier-dl}, we update the $\mL$ by solving \eqref{fp_bcd_l_m} with $\mW$ being set to
$\mW^\star$ is given by \eqref{fp_bcd_W}. This concludes the derivation of the algorithm. The overall \ac{fp}-\ac{bcd} algorithm for solving problem \eqref{eq:optim} is summarized in Algorithm \ref{Algorithm1}.

\subsubsection*{Convergence}
It is straightforward to show that the proposed algorithm converges to a stationary point. To show this, we first note that the objective in each step of the \ac{bcd} algorithm, i.e., the marginal objectives in the outer loop, is a non-decreasing function of the underlying variables. This guarantees that in each iteration, the objective is increasing. On the other hand, the downlink power constraint guarantees that the achievable weighted sum-rate is bounded from above, and therefore the iterative increments of the objective in the algorithm are limited by this upper bound. This implies that Algorithm \ref{Algorithm1} always converges to a stationary point. Noting that this point is a combination of marginal maximizers, we can further conclude that this stationary point is a local maximizer.

\subsubsection*{Complexity}
The computational complexity of Algorithm \ref{Algorithm1} is dominated by \eqref{fp_bcd_W} which is the classical operation for linear \ac{rzf}. To see this point, we first note that finding the optimal dual variables, i.e., $\omega_k$ and $q_k$, imposes computational complexity of the order ${\mathcal{O}}(K^2M)$. The inner loop, i.e., algorithm $\mA$, further is linear in all dimensions and performs a grid search over a set of $L$ grid points in each iteration. Its computational complexity is hence ${\mathcal{O}}(NMLK)$. The computation of $\mathbf{W}$ deals with matrix inversion and imposes the computational complexity of the order ${\mathcal{O}}(KM^2+M^3)$. Comparing these operations, we observe that the latter is the one which dominates the order of complexity in the overall algorithm. Noting that \eqref{fp_bcd_W} is a classical linear precoding operation, one can conclude that the proposed algorithm imposes the same order of complexity as in classical linear precoding. More precisely, Algorithm~\ref{Algorithm1} scales with system dimensions as ${\mathcal{O}}(2K^2MI+KM^2I+M^3I+MNLKI)$ with $I$ being the number of iterations in the outer loop. Our numerical investigations depict that the required number of iterations is rather small implying that the complexity of the proposed algorithm is comparable to that of classical linear precoding.

\begin{algorithm}[!t]
  \caption{\ac{fp}-\ac{bcd} Hybrid Beamforming}
  \label{Algorithm1}
  \begin{algorithmic}[1]
  \REQUIRE Weights $\lambda_k$ for $k\in [K]$
  \STATE Initialize $\epsilon$, feasible $\mW$ and $\mL$, and resolution $L$
    \REPEAT 
      \STATE Update dual variables $\omega_k$ and $q_k$  via \eqref{fp_bcd_omage_k} and \eqref{fp_bcd_q_k} for $k\in\dbc{K}$
      \STATE Computed $\mT$ and $\mU$ from the dual variables via \eqref{eq:U-T}
      \STATE Update $\mathbf{W}$ via \eqref{fp_bcd_W}
      \FOR{$m=1:M$}
      \STATE Compute $f_{m}(\ell)$ via \eqref{eq:fm}
      \FOR{$n=1:N$}
      \STATE Update $\ell_{n,m}$ via \eqref{fp_bcd_l_m}
      \ENDFOR
      \ENDFOR
    \UNTIL{fractional increase of objective falls below $\epsilon$}
    \STATE Scale $\mathbf{W}$ as ${\mW} =\sqrt{{P_\dl}/{\tr{{\mW}^{\her} {\mW}}}} {\mW}$
    \RETURN Precoding matrix $\mW$ and location matrix $\mL$
  \end{algorithmic}
\end{algorithm}

\subsection{Alternative Algorithm via Zero-Forcing}
Algorithm~\ref{Algorithm1} maintains reasonable complexity for moderate-size \acp{pas}. However, its computational burdens might be relatively high for large \acp{pas}. In this subsection, we present an alternative scheme based on \ac{zf} precoding which optimizes the location matrix for optimal effective \ac{zf}. This bypasses the need for solving the dual problem, i.e., the outer loop, allowing for optimization in a single loop. 

Let us assume that $M > K$. In this case, we can zero-force the effective downlink channel for a given $\mL$ as \cite{heath2018foundations}
\begin{align}\label{ZF_Beamformer}
\mW = \sqrt{\frac{{P_\dl}}{\tr{ \boldsymbol{\Gamma} \brc{\mL} }}}\mG^* \brc{\mL}\boldsymbol{\Gamma} \brc{\mL},
\end{align}
with $\boldsymbol{\Gamma} \brc{\mL} = \brc{\mG^\trp \brc{\mL}\mG^* \brc{\mL}  }^{-1}$. Using \eqref{ZF_Beamformer}, the achievable downlink weighted sum-rate is given by
\begin{align}\label{eq:R_new}
    \R^\dl \brc{\mW, \mL} = \sum_{k=1}^K \lambda_k \log \brc{1+\frac{P_\dl}{\sigma^2}\frac{1}{\tr{ \boldsymbol{\Gamma} \brc{\mL} } }}.
\end{align}

The joint design of $\mL$ and $\mW$ in this case reduces to maximizing the objective in \eqref{eq:R_new} against $\mL$. Noting that the logarithm function is strictly increasing, one can alternatively find the optimal $\mathbf{L}$ via the following optimization
\begin{align}
        &\min_{ \mathbf{L}} \tr{  \boldsymbol{\Gamma} \brc{\mL} }
        \subto {\bl_m\in \setL_m \txtfor m\in [M] }. \label{Optimization_location_zf}
\end{align}
Note that this optimization directly optimizes the \ac{zf} precoding matrix, and hence no \ac{bcd} loop, i.e., the outer loop in Algorithm~\ref{Algorithm1}, is needed in this case.

Similar to the \ac{fp}-\ac{bcd} scheme, this problem can also be solved via the Gauss-Seidel approach and grid search. Nonetheless, calculating the \textit{matrix inversion} in ${\boldsymbol{\Gamma} \brc{\mL}}$ at each search point is computationally intensive. To mitigate this burden, we invoke the Sherman-Morrison lemma to develop a rank-$1$ decomposition for this trace term, in the case of $M>K$. To this end, let us focus on the scalar optimization when we set $\ell_{m,n} = \ell$ and remaining locations in $\mL$ fixed: in this case, we have
\begin{subequations}
\begin{align}
\tr{ \boldsymbol{\Gamma} \brc{\mL} }  &=\tr{\brc{\tilde{\mG}_{m,n}^\trp \brc{\ell}\tilde{\mG}_{m,n}^* \brc{\ell}}^{-1}}\\
&=\tr{\brc{\tilde{\bg}_m^n\brc{\ell}\tilde{\bg}_m^{n\her}\brc{\ell}+\breve{\mG}_{m}^\trp\breve{\mG}_{m}^{*}}^{-1}},
\end{align}
\end{subequations}
where $\tilde{\mG}_{m,n} \brc{\ell}$ is defined in \eqref{eq:G_tild}, and we define $\breve{\mG}_{m}$ as ${\mG} \brc{\mL}$ missing the $m$-th column, i.e., 
\begin{align}\label{eq:breveG}
    \breve{\mG}_m =[\tilde{\bg}_1,\ldots,\tilde{\bg}_{m-1},\tilde{\bg}_{m+1},\ldots,\tilde{\bg}_{M}]^\trp.
    \end{align}
Using the Sherman-Morrison lemma, we can write
\begin{align}
& \hspace{-15mm} \brc{\tilde{\bg}_m^n\brc{\ell}\tilde{\bg}_m^{n\her}\brc{\ell}+\breve{\mG}_{m}^\trp\breve{\mG}_{m}^{*}}^{-1} \nonumber \\
&=\boldsymbol{\Gamma}_m -\frac{\boldsymbol{\Gamma}_m \tilde{\bg}_m^n\brc{\ell}\tilde{\bg}_m^{n\her}\brc{\ell}\boldsymbol{\Gamma}_m }
{1+\tilde{\bg}_m^{n\her}\brc{\ell}\boldsymbol{\Gamma}_m \tilde{\bg}_m^n\brc{\ell}},
\end{align}
where $\boldsymbol{\Gamma}_m = \brc{\breve{\mG}_{m}^\trp\breve{\mG}_{m}^{*}}^{-1}$ exists due to $M>K$. The scalar problem is hence written as
\begin{subequations}\label{Optimization_single_location_transformed_zf}
    \begin{align}
        &\max_{ \ell }  \;
        \frac{\tilde{\bg}_m^{n\her} \brc{\ell}
        \boldsymbol{\Gamma}_m^2
        \tilde{\bg}_m^n\brc{\ell}}
{1+\tilde{\bg}_m^{n\her}\brc{\ell}\boldsymbol{\Gamma}_m\tilde{\bg}_m^n\brc{\ell}}  \\ 
         &\subto  0 \leq \ell \leq L_m \txtand 
        \Delta\ell \leq \lvert\ell_{m,i}-\ell \rvert 
        \txtfor i\ne n,
    \end{align}
    \end{subequations}
which can be solved using a grid search similar to \eqref{fp_bcd_l_m}.

The algorithm is given in Algorithm~\ref{Algorithm3}. The non-decreasing shape of the objective in each iteration and its boundedness implies that this algorithm converges to a local minimum. 

\begin{algorithm}[!t]
\caption{Single-Loop Hybrid Beamforming}
\label{Algorithm3}
\begin{algorithmic}[1]
\REQUIRE Weights $\lambda_k$ for $k\in [K]$
\STATE Initialize $\epsilon$, feasible $\mW$ and $\mL$, and resolution $L$
\REPEAT 
  \FOR{$m\in[M]$ and $n\in[N]$}
      \STATE Solve problem \eqref{Optimization_single_location_transformed_zf} using grid search
    \ENDFOR
\UNTIL{fractional decrease of objective falls below $\epsilon$}
\STATE Calculate the ZF digital detector $\mathbf{W}$ using \eqref{ZF_Beamformer}
\RETURN Precoding matrix $\mW$ and location matrix $\mL$
\end{algorithmic}
\end{algorithm}

\color{black}

\section{Multiuser Detection in an Uplink PASS}\label{sec:H-UL}
We now consider uplink transmission in the \ac{pas}, where users transmit signals to the \ac{ap} over the Gaussian multiple access channel. In this case, the key task is to design a multiuser receiver, which jointly determines the receiver matrix and activated locations of pinching elements, such that the uplink throughput is optimized. Similar to downlink beamforming, this describes a hybrid design. 
However, unlike the downlink case, the optimal digital detector in the uplink case can be directly characterized via the well-known \ac{mmse} detector. This allows for more efficient algorithmic developments in this case. 

\subsection{Hybrid Design for Multiuser Detection}
Let $s_k\in \setC$ denote the encoded information signal at user $k$. We assume that the signals of different users are independent, zero-mean, and unit-variance. User $k$ scales its signal as $x_k = \sqrt{P_{\rm u}} s_k$, where $P_{\rm u} >0$ denotes the uplink power. The effective received signal at the \ac{ap} is given by
\begin{align}
    \br = \mG \brc{\mL} \bx + \bnu 
     &=\sqrt{P_{\rm u}} \sum_{k} s_k \bg_k\brc{\mL} + \bnu.
\end{align}

To estimate the transmitted signals, i.e., $s_1,\ldots, s_K$, the \ac{ap} employs a bank of $K$ linear filters in the digital domain. Let $\bmm_k \in \setC^M$ denote the receiver that estimates the signal of user $k$. Using this linear receiver, the signal of user $k$ is estimated as $\hat{s}_k = \bmm_k^\trp \br$, which
depends on both \textit{digital receiver} $\bmm_k$ and the \textit{analog receive beam} specified by $\mL$. This implies that the detection task is a \textit{hybrid} problem in which the digital components, i.e., $\bmm_k$ for $k\in [K]$, and analog degrees of freedom, i.e., $\mL$, are to be designed jointly. 

\subsection{Uplink Weighted Sum-Rate}
To quantify the quality of estimates at the \ac{ap}, various metrics can be used, e.g., \ac{mse} or achievable rates. To be consistent with the downlink performance metric, we adopt the uplink achievable weighted sum-rate. Considering the signal model, the achievable rate for user $k$ over the uplink channel is given by
\begin{align}
    R_k^\ul \brc{\bmm_k, \mL} = \log \brc{1 + \sinr_k^\ul \brc{\bmm_k, \mL}},
\end{align}
where $\sinr_k^\ul \brc{\bmm_k, \mL}$ represents the uplink \ac{sinr} at the \ac{ap} for user $k$, defined as
\begin{align}
    \sinr_k^\ul \brc{\bmm_k, \mL} = \frac{ \abs{\bmm_k^\trp \bg_k  \brc{\mL}}^2}{\displaystyle  \sum_{j\neq k} \abs{\bmm_k^\trp \bg_j \brc{\mL}}^2 + \frac{\varsigma^2}{P_\ul} \norm{\bmm_k}^2 }.
    \label{eq:SINR_k-ul}
\end{align}
The uplink weighted sum-rate is then given by
\begin{align}\label{eq:R_up}
    \R^\ul \brc{\mM, \mL} = \sum_{k} \beta_k R_k^\ul \brc{\bmm_k, \mL},
\end{align}
computed for some non-negative weights $\beta_k$. In this notation, the matrix $\mM \in \setC^{M\times K}$ is the \textit{digital receiver matrix} defined as $\mM = \dbc{\bmm_1 \; \cdots \bmm_K}$. The goal is to design $\mM$ and $\mL$, such that the achievable uplink weighted sum-rate is maximized.

\subsection{Optimal Hybrid Design for Uplink}

The per-user \ac{sinr} given in \eqref{eq:SINR_k-ul} features a generalized Rayleigh quotient. This is maximized when we set $\bmm_k$ to be the \ac{mmse} detector \cite{heath2018foundations}. %
As a result, the optimal digital detector for a given ${\mathbf{L}}$ is readily given by the corresponding \ac{mmse} matrix, i.e., 
\cite{bjornson2014optimal}
\begin{align}\label{MMSE_Detector_Second}
\mM = \mG^{*} \brc{\mL}\brc{\mG^\trp \brc{\mL}\mG^{*} \brc{\mL}+\frac{\varsigma^2}{P_\ul} \mI_K}^{-1}.
\end{align}
Substituting \eqref{MMSE_Detector_Second} into \eqref{eq:R_up}, we can represent $R_k^\ul \brc{\bmm_k, \mL} $ as given in \eqref{eq:WR_up_1} at the top of the next page.
\begin{figure*}
  \begin{align}
R_k^\ul \brc{\bmm_k, \mL} = \log \brc{\det\brc{\mI_M + \bg_k\brc{\mL}\bg_k^\her\brc{\mL}\brc{\sum_{j\ne k}\bg_j\brc{\mL}\bg_j^\her\brc{\mL}+\frac{\varsigma^2}{P_\ul} \mI_M}^{-1}}}\label{eq:WR_up_1}
\end{align}  
\hrule
\end{figure*}

Using Sylvester's determinant identity, the weighted sum-rate can be rewritten as
\begin{align}
    &\R^\ul \brc{\mM, \mL}
    =\beta_{\sum} \log\det\brc{\mI_K + \frac{P_\ul}{\varsigma^2}\mG^\her\brc{\mL}\mG\brc{\mL}} \nonumber \\
    &-\sum_{k=1}^{K}\beta_k\log\det\brc{\mI_{K-1} + \frac{P_\ul}{\varsigma^2}\mG_k^\her\brc{\mL}\mG_k\brc{\mL}}\label{uplink_sum_rate_not_simplified},
\end{align}
where $\beta_{\sum} = \sum_k \beta_k$, 
and $\mG_k\brc{\mL}\in \setC^{M\times K-1}$ is a reduced form of $\mG\brc{\mL}$ with its $k$-th column missing. 
Noting that the resulting weighted sum-rate only depends on $\mL$, we can write the throughput optimization problem as 
\begin{align}
    &\max_{\mL} \R^\ul \brc{ \mL}  \label{eq:optim-up}
    \subto \bl_m \in \setL_m \txtfor  m\in[M], 
\end{align}
with $\setL_m$ defined in \eqref{setL}, where we drop $\mM$ for compactness. Similar to Algorithms~\ref{Algorithm1} and \ref{Algorithm3}, we can tackle the problem in \eqref{eq:optim-up} via the Gauss-Seidel approach. Due to the similarity, we skip the detailed derivation for this case. 
By omitting terms irrelevant with $\ell_{m,n}$, we equivalently simplify to the following:
\begin{subequations}\label{Optimization_single_location_simplified}
    \begin{align}
    &\max_{ \ell } \; \beta_{\sum} \log\brc{1+\frac{P_\ul}{\varsigma^2}\tilde{\bg}_{m}^{n \trp} \brc{\ell } \tilde{\boldsymbol{\Gamma}}_m \tilde{\bg}_{m}^{n*} \brc{\ell }}\nonumber\\
    &\qquad -\sum_{k } \beta_k\log\brc{1+\frac{P_\ul}{\varsigma^2} {\tilde{\bg}_{m \backslash k }^{n\trp} \brc{\ell }} \tilde{\boldsymbol{\Gamma}}_{m\backslash k} \tilde{\bg}_{m\backslash k}^{*} \brc{\ell } }\\
        &\subto  0 \leq \ell \leq L_m \txtand 
        \Delta\ell \leq \lvert\ell_{m,i}-\ell \rvert 
        \txtfor i\ne n,
    \end{align}
    \end{subequations}
where we define
\begin{subequations}
    \begin{align*}
       \tilde{\boldsymbol{\Gamma}}_m &= \brc{\mI_K + \frac{P_\ul}{\varsigma^2}\breve{\mG}_{m}^\her\breve{\mG}_{m}}^{-1}\\
       \tilde{\boldsymbol{\Gamma}}_{m\backslash k} &= \brc{\mI_K + \frac{P_\ul}{\varsigma^2}\breve{\mG}_{m\backslash k}^\her\breve{\mG}_{m\backslash k}}^{-1}
    \end{align*}
\end{subequations}
with $\breve{\mG}_{m}$ defined in \eqref{eq:breveG} and $\breve{\mG}_{m\backslash k}$ being its reduced version whose $k$-th column is dropped. Similarly, ${\tilde{\bg}_{m \backslash k }^{n} \brc{\ell }}$ denotes ${\tilde{\bg}_{m }^{n} \brc{\ell }}$ whose $k$-th entry is dropped. As in Algorithms~\ref{Algorithm1} and \ref{Algorithm3}, this problem can be solved efficiently by grid search. The final algorithm is presented in Algorithm~\ref{Algorithm2}.

\begin{algorithm}[!t]
\caption{Greedy Hybrid Receiver}
\label{Algorithm2}
\begin{algorithmic}[1]
\REQUIRE Weights $\beta_k$ for $k\in [K]$
\STATE Initialize $\epsilon$, feasible $\mW$ and $\mL$, and resolution $L$
\REPEAT 
  \FOR{$m\in[M]$ and $n\in[N]$}
      \STATE Solve problem \eqref{Optimization_single_location_simplified} using the grid search
    \ENDFOR
\UNTIL{fractional increase of objective in \eqref{eq:optim-up} falls below $\epsilon$}
\STATE Determine optimal digital detector $\mathbf{M}$ using \eqref{MMSE_Detector_Second}
\RETURN Receiver matrix $\mM$ and location matrix $\mL$
\end{algorithmic}
\end{algorithm}

\begin{figure*}[t]
\centering
    \subfigbottomskip=2pt
	\subfigcapskip=0pt
\setlength{\abovecaptionskip}{4pt}
   \subfigure[Downlink]
    {
        \includegraphics[height=0.28\textwidth]{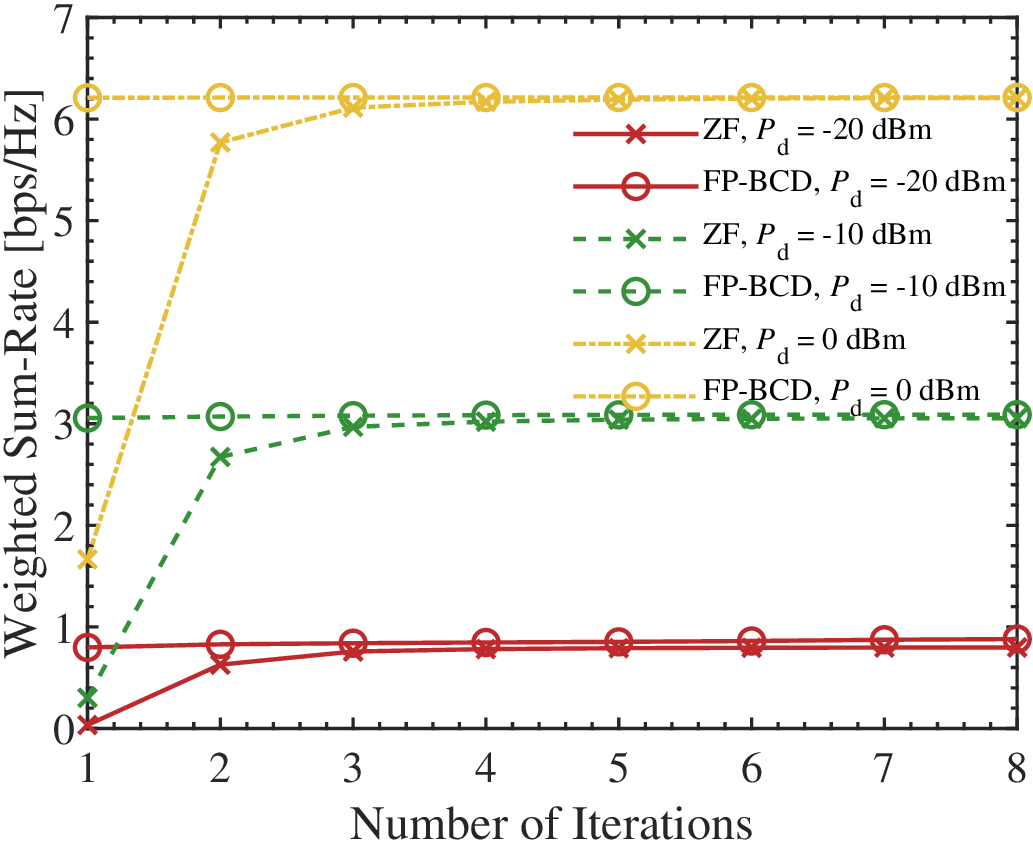}
	   \label{Figure: Convergence2}	
    }\hspace{60pt}
    \subfigure[Uplink]
    {
        \includegraphics[height=0.28\textwidth]{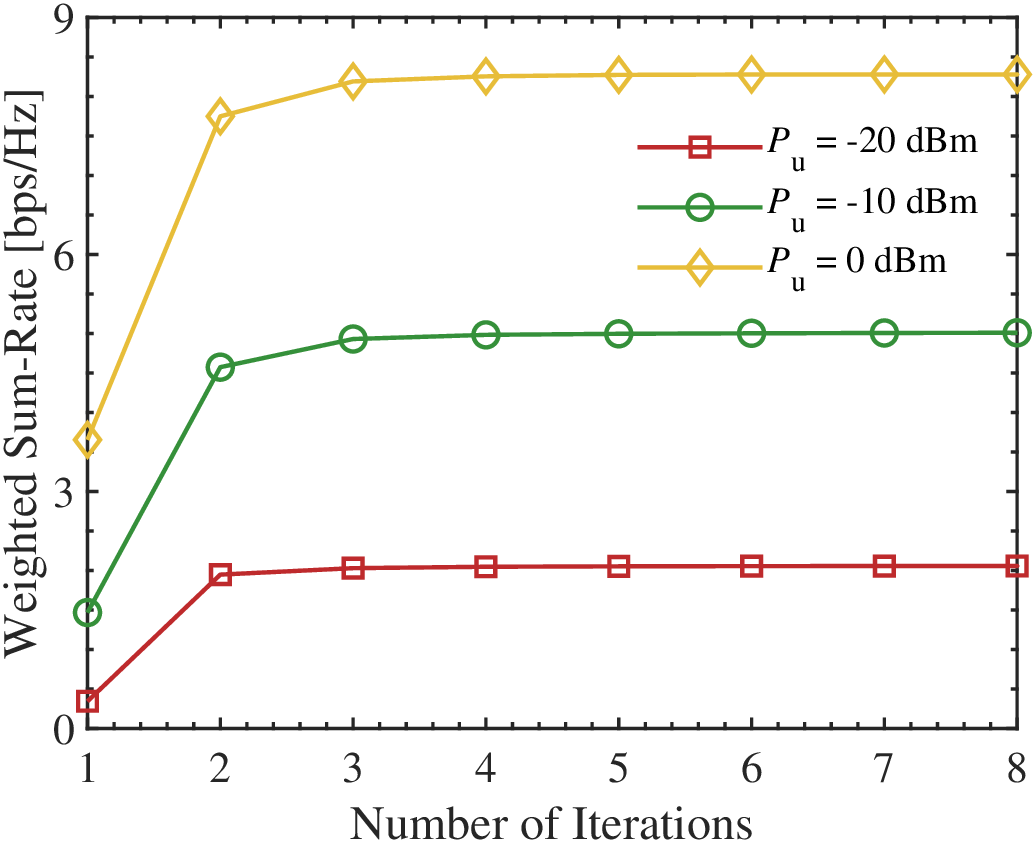}
	   \label{Figure: Convergence3}	
    }
\caption{Convergence of the proposed iterative algorithms for MIMO-PASS uplink and downlink transmission.}
\label{Figure: Convergence}
\vspace{-10pt}
\end{figure*}

\section{Numerical Investigations}
\label{sec:Numerical}
We now validate the proposed algorithms through numerical experiments. Through the numerical simulations, we investigate \acp{pas} and the proposed algorithms in two major aspects: ($i$) we evaluate the convergence and precision of the proposed algorithms. ($ii$) We compare \ac{mimo}-\acp{pas} against classical \ac{mimo} baselines in the same simulated environment to assess the order of gain achieved by this technology. 

\subsection{Experimental Setting}

Consider a rectangular region with side lengths $D_x$ and $D_y$ along the $x$- and $y$-axes, respectively, centered at $\left[D_x/2,D_y/2,0\right]$. Within this region, $K$ single-antenna users are randomly and uniformly distributed. A schematic illustration of this setup is provided in {\figurename} {\ref{Fig1_Schematic}}. The $M$ waveguides are assumed to be positioned at a height of $a=5$ m and cover the entire area. These waveguides are spaced uniformly along the $y$-axis with a separation distance of $d={D_y}/(M-1)$. Each waveguide has a length of $D_x$, i.e., $L_m = D_x$ for $m\in[M]$, and the antenna elements are spaced at least half a wavelength apart, i.e., $\Delta \ell = \lambda/2$. For the simulations, the following parameters are adopted unless otherwise specified: we consider $M=5$ waveguides, each equipped with $N=6$ pinching elements and fed by a dedicated RF chain, serving $K=4$ users. The side lengths of the rectangular region are set to $D_x=50$ m and $D_y=6$ m. The noise variance in both uplink and downlink is $\sigma^2=\varsigma^2=-90$ dBm. The carrier frequency is $f=28$ GHz, and the refractive index is $i_{\mathrm{ref}}=1.44$ \cite{ding2024flexible}. We use uniform weights, i.e., $\lambda_k = \beta_k = 1/K$, in both uplink and downlink. Furthermore, the path-loss is computed via the free-space propagation model, i.e., $\alpha_{k}=1$ for $k\in[K]$. Numerical results are obtained by averaging $500$ random seeds.

For implementation of the proposed algorithms, i.e., Algorithms~\ref{Algorithm1}, \ref{Algorithm3}, and \ref{Algorithm2}, the predefined threshold is set to $\epsilon=10^{-3}$. For the one-dimensional grid search, the grid resolution is set to $L=10^5$, unless stated otherwise. Since the performance of the FP-BCD algorithms may depend on the initial parameters, the digital precoding matrix and location matrix in Algorithm~\ref{Algorithm1} are initialized using the results obtained from the \ac{zf}-based algorithm, i.e., Algorithm~\ref{Algorithm3}. For the \ac{zf}-based beamforming and \ac{mmse}-based detection designs, the locations of the pinching elements are initialized randomly.

\subsection{Baseline Architectures}
We compare a \ac{mimo}-\ac{pas} with $M$ waveguides each~having $N$ pinching elements against three benchmark technologies: \textit{conventional \ac{mimo}}, \textit{massive \ac{mimo} (mMIMO)}, and \textit{hybrid beamforming-based \ac{mimo} (hMIMO)}. To ensure a fair comparison, the following configurations are considered for the baselines:
\begin{inparaenum}
    \item[(1)] The \textit{conventional \ac{mimo}} system is equipped with $M$ antennas, each connected to a dedicated RF chain, and employs fully digital signal processing. Specifically, FP-based digital precoding is used in the downlink, and MMSE-based digital detection is applied in the uplink. From an implementation perspective, this setup incurs approximately the same RF cost as the proposed PASS, as it utilizes the same number of RF chains at the AP.        
    \item[(2)] The \textit{mMIMO} system features an array of $MN$ antennas, each with its own RF chain, and employs fully digital signal processing for both uplink and downlink. The beamforming for downlink and uplink is designed using FP-based and MMSE-based methods, respectively. Due to its large number of RF chains, this baseline represents a significantly more expensive design compared to both conventional MIMO and the proposed PASS.
    \item[(3)] The \textit{hMIMO} system utilizes a hybrid transceiver with $M$ RF chains, each connected to $N$ antenna elements via a network of tunable phase shifters. It employs hybrid analog-digital signal processing for both uplink and downlink, which is designed using the algorithms in \cite{yu2016alternating}. While this setup imposes the same RF cost as the proposed PASS, it may suffer from poor power efficiency due to high power splitting losses.
\end{inparaenum}
In simulations, all baselines are based on a half-wavelength spaced uniform linear array centered within the square region at $[D_x/2, D_y/2, a]$, and aligned along the $y$-axis.

\subsection{Numerical Results}

\begin{figure*}[!t]
\centering
    \subfigbottomskip=2pt
	\subfigcapskip=0pt
\setlength{\abovecaptionskip}{4pt}
    \subfigure[Downlink. $P_\dl=0$ dBm]
    {
        \includegraphics[height=0.28\textwidth]{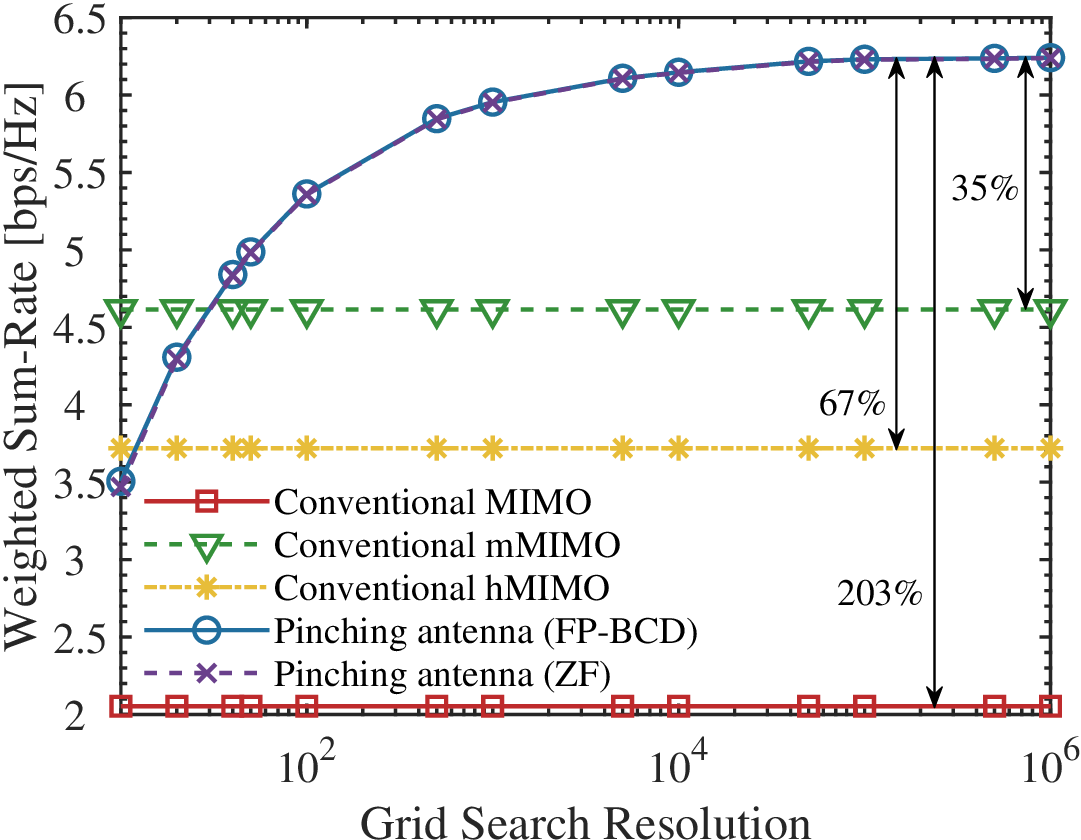}
	   \label{Figure: DL_Sum_Rate_Precision}	
    }\hspace{60pt}
   \subfigure[Uplink. $P_\ul=0$ dBm]
    {
        \includegraphics[height=0.28\textwidth]{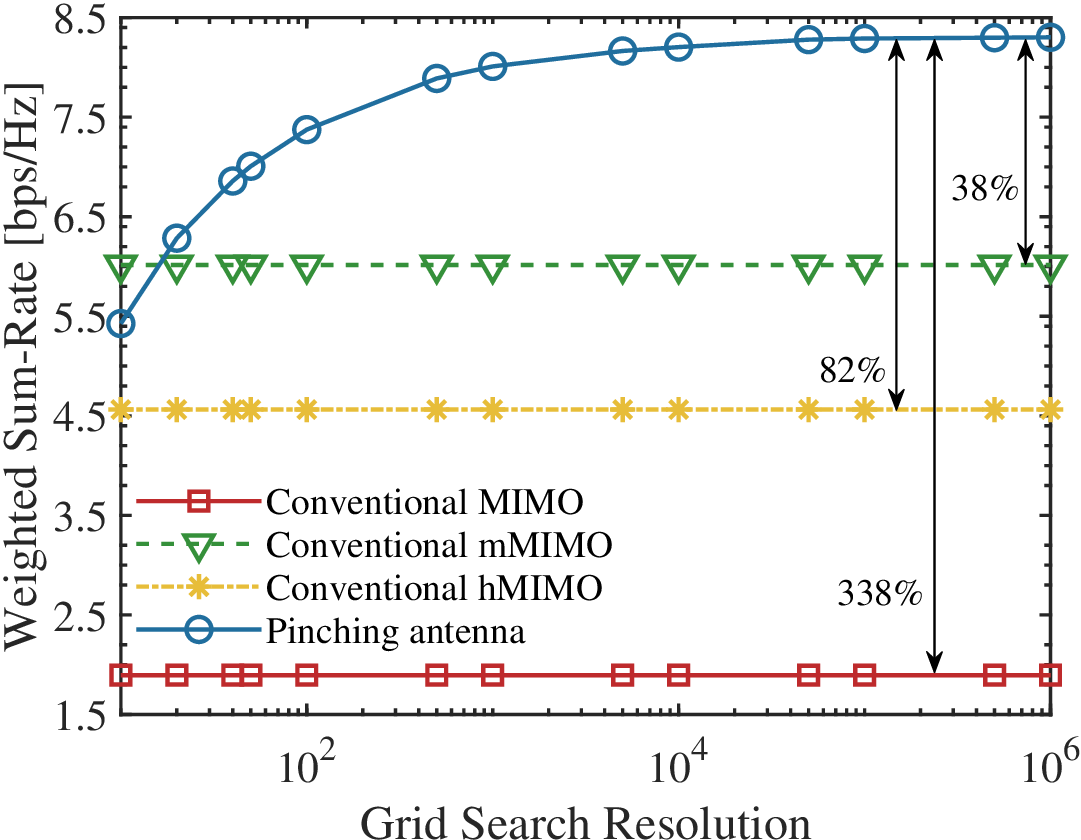}
	   \label{Figure: UL_Sum_Rate_Precision}	
    }
\caption{Weighted sum-rate vs the grid search resolution. Arrows indicate the gain achieved by MIMO-PASS over the baselines.}
\label{Figure: Sum_Rate_Precision}
\vspace{-10pt}
\end{figure*}

\subsubsection*{Convergence.}
{\figurename}~{\ref{Figure: Convergence}} illustrates the convergence behavior of the proposed algorithms for different values of transmitted power. Specifically, {\figurename} {\ref{Figure: Convergence2}} shows the downlink weighted sum-rate achieved by the FP-BCD and ZF-based methods as a function of the number of iterations. As observed, the weighted sum-rate gradually increases with the number of iterations and eventually converges to a stable value for all considered values of $P_{\rm{d}}$, which confirms the effectiveness of the proposed algorithms. Furthermore, increasing the transmit power improves the system throughput, i.e., the weighted sum-rate. Notably, the weighted sum-rate at the initial point of the FP-BCD method is the same as that at the converged point of the ZF-based method, as the FP-BCD method is initialized using the optimized results from the ZF-based method. {\figurename} {\ref{Figure: Convergence3}} demonstrates the convergence behavior of the MMSE-based algorithm for optimizing the uplink weighted sum-rate. Under the considered setup, the proposed algorithms converge within $5$ iterations, verifying their fast convergence.

\subsubsection*{Grid Search Resolution.}
As the one-dimensional grid search is utilized in Algorithms~\ref{Algorithm1}, \ref{Algorithm3}, and \ref{Algorithm2}, the achievable uplink and downlink throughput is influenced by the grid resolution, i.e., the number of discrete points along the waveguide, $L$. To illustrate this impact, {\figurename} {\ref{Figure: DL_Sum_Rate_Precision}} and {\figurename} {\ref{Figure: UL_Sum_Rate_Precision}} plot the downlink and uplink weighted sum-rates achieved by the proposed algorithms as a function of the grid search resolution $L$, respectively. As shown in these graphs, the weighted sum-rates for both downlink and uplink increase monotonically with the grid search resolution. Additionally, we observe that the achieved sum-rates saturate when $L$ reaches approximately $10^5$. This suggests that setting the grid resolution to $10^5$ nearly attains the performance upper bound of the PASS.

For a thorough study, we also compare the weighted sum-rate achieved by the MIMO-PASS with those achieved by the baseline fixed-location antenna systems, i.e., \emph{MIMO}, \emph{mMIMO}, and \emph{hMIMO}. As seen in both {\figurename} {\ref{Figure: DL_Sum_Rate_Precision}} and {\figurename} {\ref{Figure: UL_Sum_Rate_Precision}}, the MIMO-PASS achieves significantly better weighted sum-rate performance compared to these baselines. Notably, even when compared to a fully digital massive MIMO system with the same number of antennas as the PASS, the PASS achieves a gain of more than $30\%$. This gain further increases to over $200\%$ when compared to small-scale fully digital MIMO systems. Such promising improvements are primarily attributed to the capability of \acp{pas} to flexibly tune pinching locations on waveguides, enabling collaborative establishment of strong and stable LoS paths, i.e., a \emph{nearly-wired} transmission link.

\begin{figure*}[!t]
\centering
    \subfigbottomskip=2pt
	\subfigcapskip=0pt
\setlength{\abovecaptionskip}{4pt}
    \subfigure[Transmit power $P_\dl$]
    {
        \includegraphics[height=0.25\textwidth]{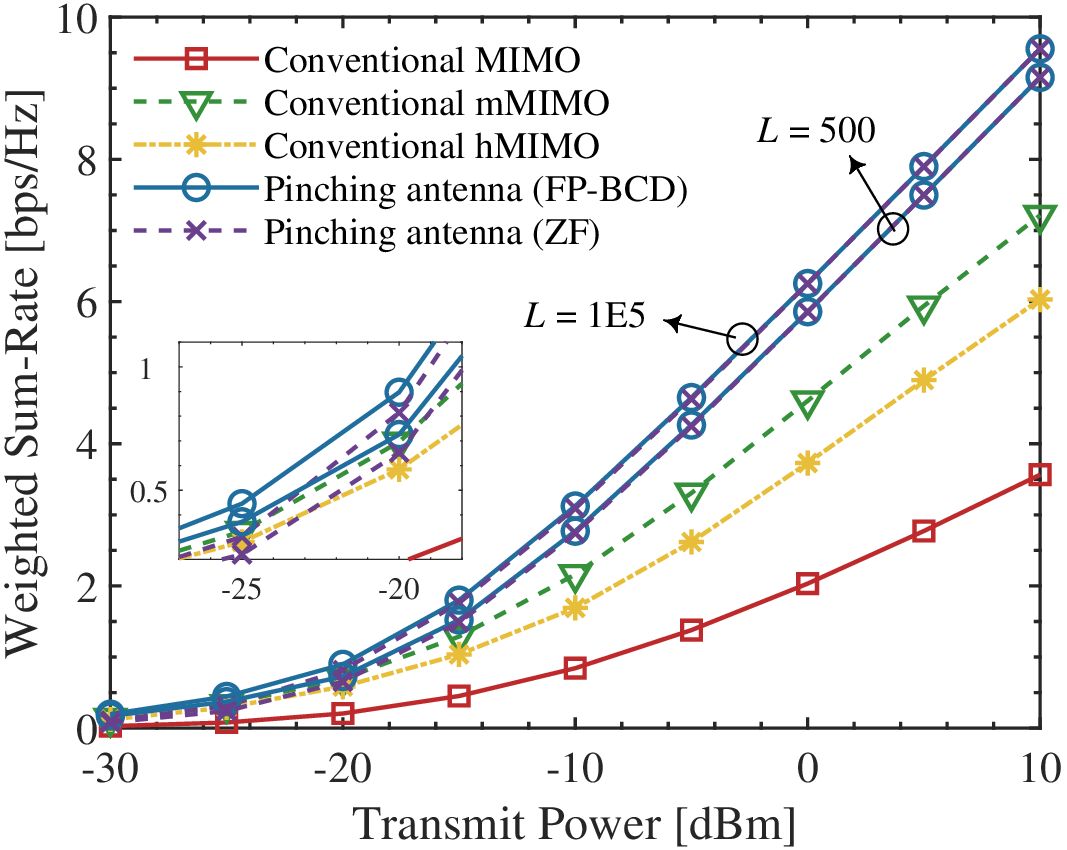}
	   \label{Figure: DL_Sum_Rate_Power}	
    }
   \subfigure[Side length $D_x$. $P_\dl=0$ dBm]
    {
        \includegraphics[height=0.25\textwidth]{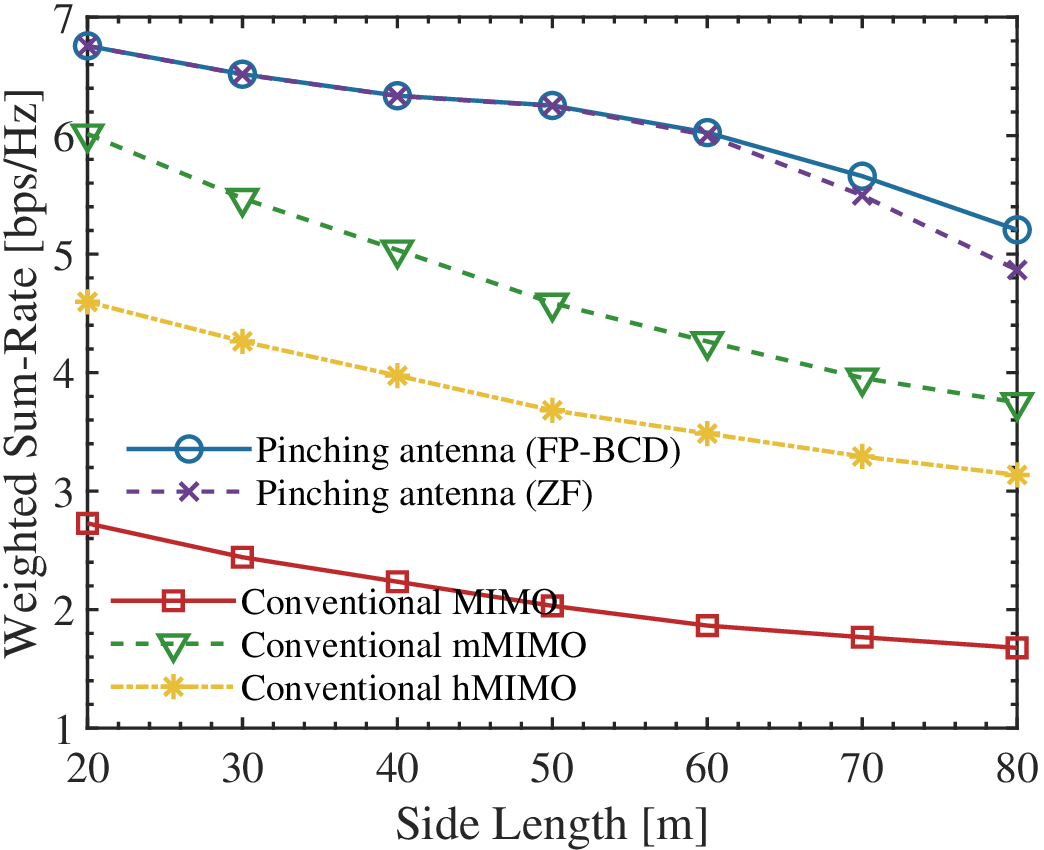}
	   \label{Figure: DL_Sum_Rate_Distance}	
    }
    \subfigure[Number of PAs $N$. $P_\dl=0$ dBm]
    {
        \includegraphics[height=0.25\textwidth]{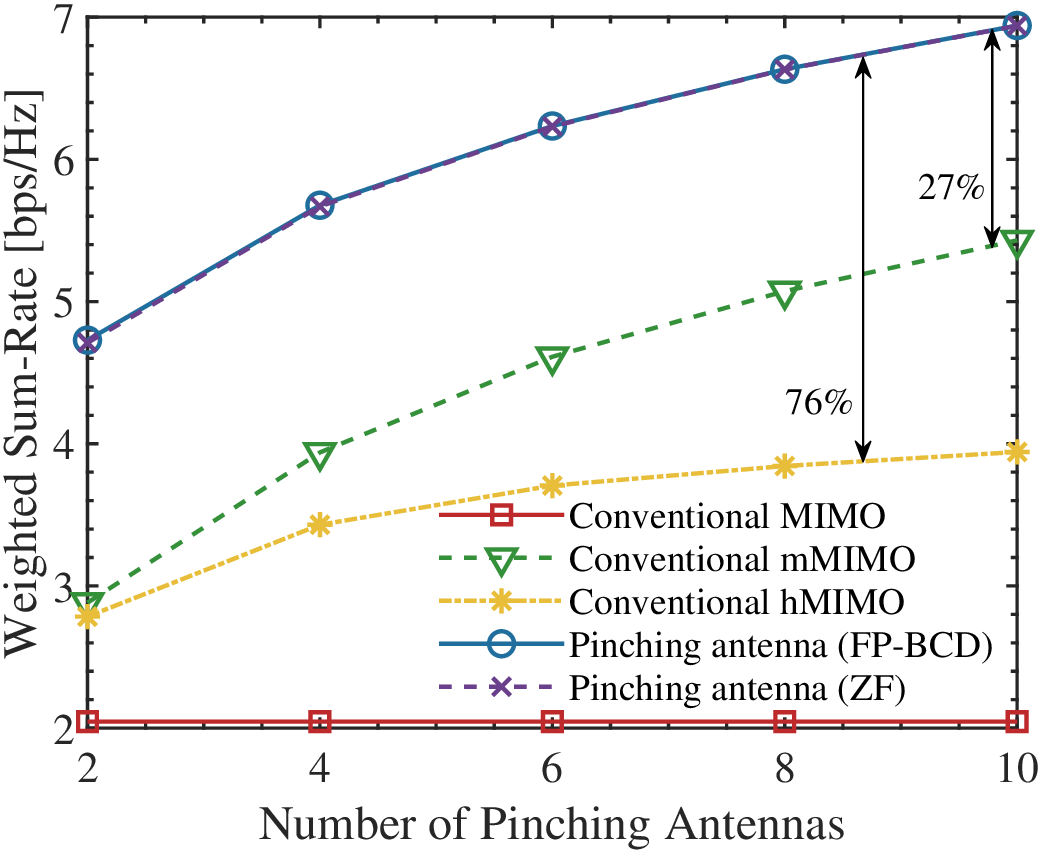}
	   \label{Figure: DL_Sum_Rate_Antenna}	
    }
\caption{Achievable downlink weighted sum-rate: arrows indicate the gain achieved by MIMO-PASS over the baselines.}
\label{Figure: DL_Sum_Rate}
\vspace{-10pt}
\end{figure*}

\subsubsection*{Downlink Weighted Sum-Rate.}
{\figurename} {\ref{Figure: DL_Sum_Rate_Power}} shows the downlink weighted sum-rate as a function of the maximum total transmit power $P_\dl$, which compares performance of the MIMO-PASS with the conventional fixed-location antenna systems. The results indicate that the sum-rate increases with $P_\dl$ for all schemes. Notably, the PASS achieves significantly higher throughput than the conventional systems, particularly in the medium- and high-power regimes. From the graph, we observe that increasing the grid search resolution, $L$, improves the weighted sum-rate achieved by the PASS, which is consistent with the what we observed in {\figurename} {\ref{Figure: DL_Sum_Rate_Precision}}. Another important observation is that in the medium- and high-power regimes, the ZF-based method achieves nearly the same performance as the FP-BCD method. However, in the low-power regime, the FP-BCD method enables the PASS to achieve a higher sum-rate than the ZF-based method. This is expected, as ZF beamforming is asymptotically optimal in the high \ac{snr} regime \cite{bjornson2014optimal,heath2018foundations}. This also explains the behavior observed in {\figurename} {\ref{Figure: Convergence2}} and {\figurename} {\ref{Figure: DL_Sum_Rate_Precision}}, where the ZF-based method achieves nearly the same performance as the FP-BCD method.

{\figurename} {\ref{Figure: DL_Sum_Rate_Distance}} illustrates the sum-rate as a function of the side length of the square region along the $x$-axis, $D_x$, while $D_y$ is fixed. As the side length increases, the performance gain of the PASS over the conventional baseline improves significantly. This is because a larger side length increases the average distance between users and the center of the region, resulting in higher path loss for the conventional systems. In contrast, the PASS can flexibly position its radiating elements closer to the users, which effectively enhances the system throughput. Furthermore, as shown in the graph, as $D_x$ increases, the FP-BCD method outperforms the ZF-based method. This is because a larger side length leads to more severe path loss, which creates a low-SNR scenario where ZF performs poorly.

{\figurename} {\ref{Figure: DL_Sum_Rate_Antenna}} further depicts the weighted sum-rate as a function of the number of pinching elements along each waveguide, $N$. As shown in the graph, the sum-rates achieved by both the MIMO-PASS setting and the baseline mMIMO and hMIMO schemes increase monotonically with the number of pinching elements. This is due to the increased spatial DoFs and array gains provided by adding more antennas. The graph also highlights the weighted sum-rate performance gain of the PASS over the baseline systems. Specifically, tuning the locations of the pinching elements introduces sum-rate improvements of $27\%$ and $76\%$ compared to the mMIMO and hMIMO schemes, respectively. These results, along those in {\figurename} {\ref{Figure: DL_Sum_Rate_Power}} and {\figurename}~{\ref{Figure: DL_Sum_Rate_Distance}}, validate the effectiveness of the MIMO-PASS in enhancing the throughput in downlink transmission.

\subsubsection*{Uplink Weighted Sum-Rate.}
\begin{figure*}[!t]
\centering
    \subfigbottomskip=2pt
	\subfigcapskip=0pt
\setlength{\abovecaptionskip}{5pt}
    \subfigure[Transmit power $P_\ul$]
    {
        \includegraphics[height=0.28\textwidth]{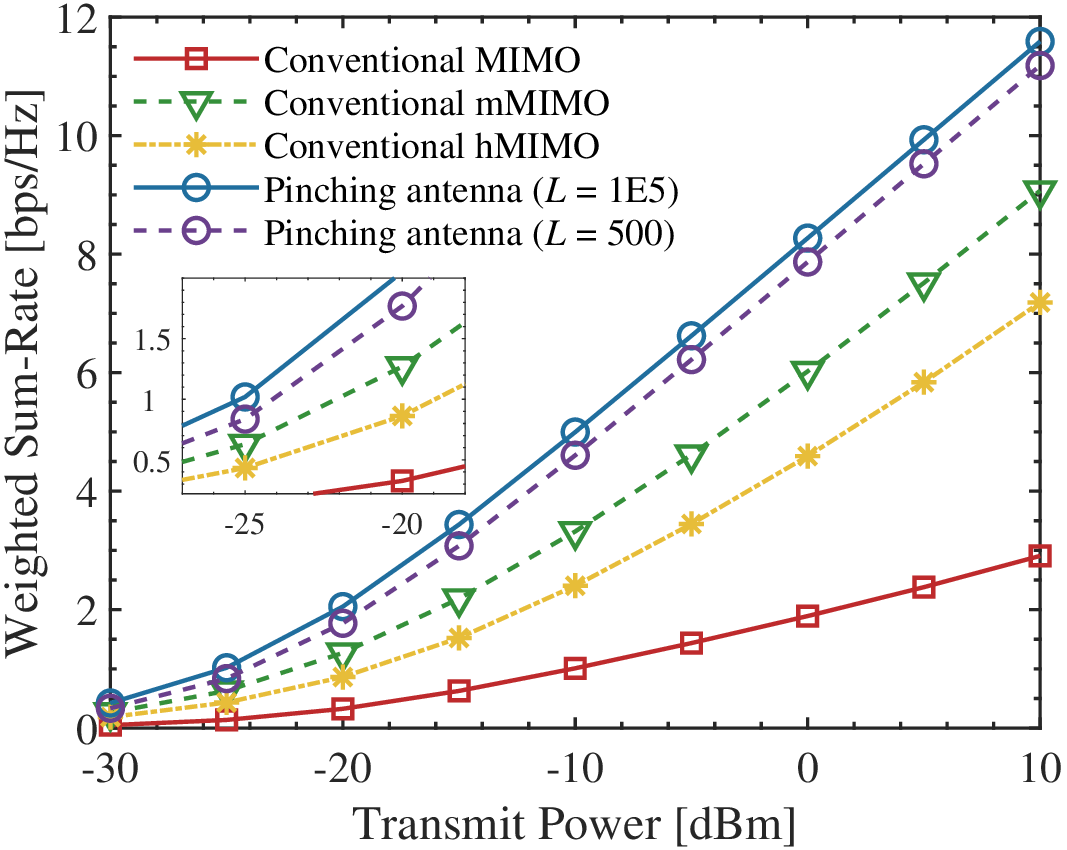}
	   \label{Figure: UL_Sum_Rate_Power}	
    }\hspace{60pt}
   \subfigure[Number of users $K$. $P_\ul=0$ dBm]
    {
        \includegraphics[height=0.28\textwidth]{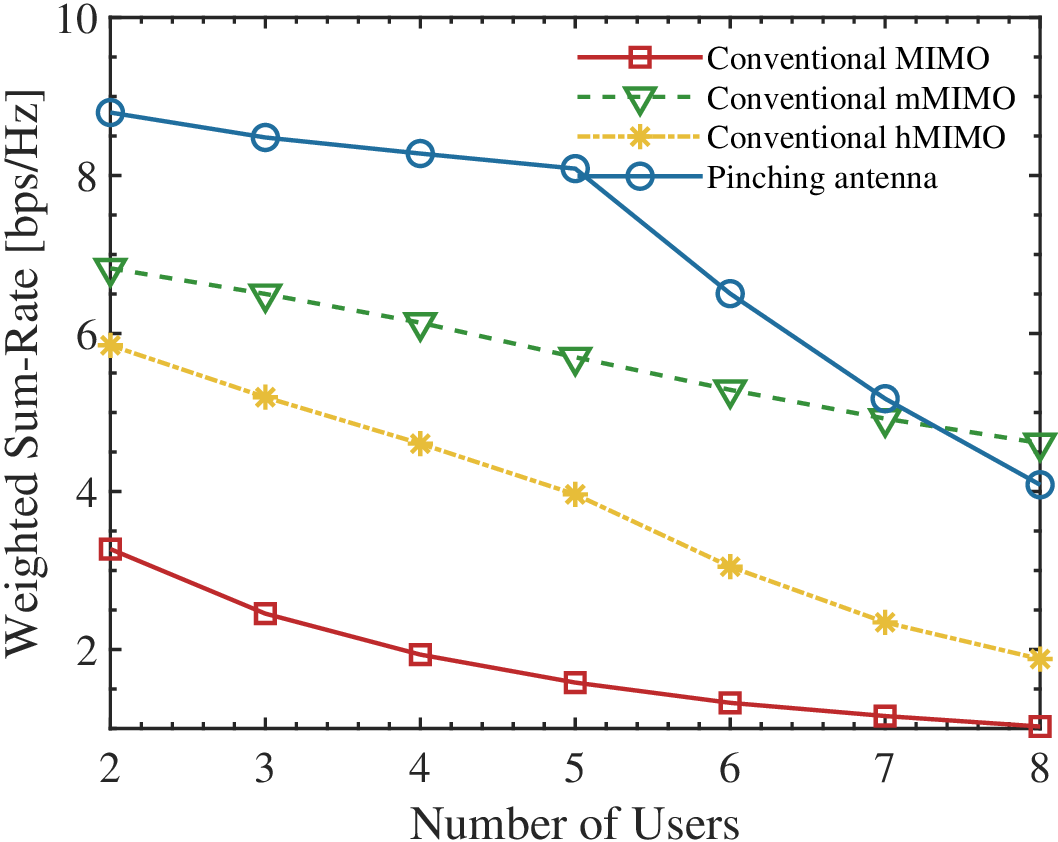}
	   \label{Figure: UL_Sum_Rate_User_Number}	
    }
\caption{Weighted uplink sum-rate achieved by MIMO-PASS against the uplink transmit power and number of users.}
\label{Figure: UL_Sum_Rate}
\vspace{-7pt}
\end{figure*}

{\figurename} {\ref{Figure: UL_Sum_Rate_Power}} depicts the achievable uplink weighted sum-rate as a function of the per-user transmit power $P_\ul$. It is observed that, across the considered range of transmit powers, the proposed MMSE-based method effectively optimizes the locations of the pinching elements, enabling the PASS to achieve higher throughput than conventional fixed-location antenna systems. Additionally, we observe that increasing the grid search resolution $L$ improves the weighted sum-rate achieved by the MIMO-PASS. This is consistent with the earlier results in {\figurename} {\ref{Figure: UL_Sum_Rate_Precision}}.

{\figurename} {\ref{Figure: UL_Sum_Rate_User_Number}} further depicts the weighted sum-rate as a function of the number of users, $K$. The sum-rates achieved by both the MIMO-PASS and the baseline systems decrease monotonically as the number of users increases, primarily due to the rise in inter-user interference. As shown in the graph, when the number of users does not exceed the number of waveguides or RF chains, i.e., $K\leq M=5$, the PASS achieves a higher sum-rate than the baseline schemes. However, once $K>M$, the sum-rate achieved by the PASS decays rapidly. This is because the PASS is essentially a hybrid beamforming system. When the number of users exceeds the number of RF chains, the system cannot provide sufficient DoFs to effectively mitigate inter-user interference. This explains why the fully digital mMIMO system outperforms the MIMO-PASS when $M=8$. Nevertheless, compared to fixed-location antenna systems with the same number of RF chains, i.e., conventional MIMO and hMIMO, the MIMO-PASS still can significantly enhance the throughput due to its ability to flexibly tune the locations of antennas and establish a strong LoS link.

\section{Conclusion}
\label{sec:Conc}
We have studied uplink and downlink transmission in a multiuser \ac{mimo}-\ac{pas} system, in which the \ac{ap} deploys an array of pinched waveguides at its front-end. Invoking the \ac{fp} framework, \ac{bcd} scheme and Gauss-Seidel approach, several low-complexity algorithms for hybrid beamforming in downlink and detection in uplink were developed. These algorithms solve a \textit{hybrid} problem, where the digital units, i.e., precoding or receiver matrix, is optimized jointly with  
the activated locations of the pinching elements on the waveguides. 
The effectiveness of the proposed algorithms was validated through several numerical experiments. The results of this study showcase the significant superiority of \ac{mimo}-\ac{pas} over conventional fixed-location designs including both massive \ac{mimo} and the benchmark hybrid analog-digital architectures. Our results highlight the potential of \acp{pas} as a key reconfigurable antenna system technology for next-generation wireless systems urging for further research in this direction.

\bibliographystyle{IEEEtran}
\bibliography{icme2022template}
\end{document}